\documentclass[aps,pre,twocolumn,groupedaddress,showpacs]{revtex4}

\usepackage{epsfig,amsmath,amsbsy,amssymb}

\usepackage{graphicx}
\usepackage{bm}

\usepackage[dvipdfm, pdftitle={text.pdf},
       pdfauthor={Bjoern Zelinski, Nina Mueller, Jan Kierfeld},
      pdfkeywords={Dynamics of single microtubules under force}
       ]{hyperref}

\usepackage{color,colordvi}
\definecolor{Red}{rgb}{0.9,0.0,0.1}

\begin{document}

\title{Dynamics and length distribution 
  of microtubules under force and confinement}

\author{Bj\"orn Zelinski\footnote{bjoern.zelinski@tu-dortmund.de}, 
Nina M\"uller\footnote{nina2.mueller@tu-dortmund.de} 
and Jan Kierfeld\footnote{jan.kierfeld@tu-dortmund.de}}
\affiliation{Physics Department, TU Dortmund University, 
44221 Dortmund, Germany}

\date{\today}
\begin{abstract}
We investigate the microtubule polymerization 
dynamics  with catastrophe and rescue events 
for  three different confinement scenarios, which mimic typical
cellular environments: 
(i)  The microtubule  is confined by  rigid and fixed   walls, 
(ii) it grows under constant force,  and 
(iii) it grows against an elastic obstacle with a linearly 
  increasing force. 
We use realistic catastrophe models and analyze the 
microtubule dynamics, the resulting microtubule 
length distributions, and force generation by stochastic 
and mean field calculations; 
in addition, we perform stochastic simulations. 
Freely growing microtubules exhibit a phase of  
bounded growth with finite microtubule length and 
a phase of unbounded growth. 
The main results for the three confinement scenarios are as follows: 
(i) In confinement by fixed rigid walls, we find 
    exponentially decreasing or increasing stationary
    microtubule length distributions instead of bounded or 
    unbounded phases, respectively.  
   We introduce a realistic model for wall-induced catastrophes
  and investigate the behavior of the average length 
  as a function of microtubule growth parameters.
(ii) Under a  constant force  the boundary
   between bounded and unbounded growth  is shifted  to higher tubulin
   concentrations and rescue rates. The critical 
  force $f_c$ for the  transition from unbounded to bounded growth
   increases logarithmically with  tubulin
   concentration and the rescue rate, and it is smaller than the stall force.
(iii) For microtubule  growth against an elastic obstacle, 
  the microtubule length and polymerization force can be regulated 
  by microtubule growth parameters.  For 
  zero rescue rate, we find that the average polymerization force depends
  logarithmically on the tubulin concentration and is always smaller than the
  stall force in the absence of catastrophes and rescues. 
  For a  non-zero  rescue rate, we find a sharply peaked steady-state 
  length distribution, which is tightly controlled by 
  microtubule growth parameters. The corresponding 
  average microtubule length self-organizes such that the  
  average polymerization force equals the critical force $f_c$ for the  
  transition from unbounded to bounded growth. 
  We also 
  investigate the force dynamics if growth parameters are perturbed 
  in dilution experiments. 
  Finally, we show the robustness of our results 
  against changes of catastrophe 
  models and load distribution factors. 
\end{abstract}


\maketitle

\section{Introduction}

Microtubules (MTs) are  one of the
main components of the cytoskeleton in eukaryotic cells. 
Their static and 
dynamic properties are essential for many cellular processes. MTs serve
as pathways for molecular motor proteins \cite{Vale1987} and contribute to
cell stiffness \cite{How2001}. 
Dynamic MTs play a crucial role in the constant 
reorganization of the cytoskeleton, and 
single MTs can generate polymerization forces up to several
pN \cite{DY97}. These forces are used 
in various intracellular positioning processes \cite{D05}, 
such as  positioning of 
 the cell nucleus \cite{daga2006} or chromosomes during
mitosis , establishing cell polarity \cite{SD07}, or regulation of
cell shapes \cite{Picone2010,Dehmelt2003}. 
 In many  cellular processes, MTs
establish and maintain a characteristic length in response to 
forces exerted, for example, from the confining  cell cortex
\cite{Picone2010}.

The fast spatial reorganization of MTs 
 is based on the dynamic instability:
Polymerization phases are stochastically interrupted by catastrophes
which initiate phases of fast depolymerization;
fast depolymerization terminates stochastically in a 
 rescue event followed again by a polymerization  phase
\cite{Mitch1984}.  
This complex dynamic behavior with catastrophes and rescue events 
is central to a rapid remodelling of MTs in the cytoskeleton, 
but it also affects their ability to generate  polymerization forces.  
We will show that, in general,  the dynamic instability {\em decreases} the 
average polymerization force of a single MT.

In this article we theoretically 
investigate the polymerization dynamics of a single
 MT under force or confinement 
and in the presence of the MT  dynamic instability.  We
use a coarse-grained polymerization model with dynamic instability and
characterize spatial and temporal behavior in three different scenarios,
which mimic typical cellular environments that  can also be reproduced 
 \textit{in vitro}:
(i) {\em  Confinement:} The MT is confined  between
 fixed  rigid walls, which cannot be deformed by the microtubule. 
Such confinement is realized in fixed solid chambers \cite{Faiv2008}.
(ii) {\em Constant force:} 
 A constant force is acting on the MT. 
Constant forces can be realized by optical tweezers with a force clamp 
control \cite{Schek2007}.
(iii) {\em Elastic obstacle:}
The microtubule grows against an elastic obstacle, which resists
further growth by a force growing linearly with displacement.
Elastic forces can be realized by optical tweezers without force clamp
\cite{Kers2006,Laan2008}. 
For all three confinement scenarios (i)--(iii) we focus on the resulting 
length distributions of MTs and for scenarios (ii) and (iii), 
we calculate the  polymerization force that a single 
MT can  generate.

Dynamic MTs also initiate regulation processes or are subject to 
regulation.  Dynamic 
MTs can  activate or
deactivate proteins upon contacting  
the cell membrane \cite{storer1999}, or they can activate actin polymerization within 
the cell cortex \cite{Mitch1988,Rod2003}.  At the same time, polymerizing MTs 
 are also targets of cellular
regulation mechanisms \cite{Athale2008}, which affect their 
dynamic properties.

The dynamic instability of MTs  
enables various  regulation mechanisms of  MT dynamics.
Catastrophes and rescues result from the  hydrolysis
 of GTP-tubulin within MTs. 
When GTP-tubulin is incorporated
into the tip of a growing MT, it forms a stabilizing GTP-cap. The loss of this
GTP-cap due to hydrolysis of GTP-tubulin to GDP-tubulin, causes a
catastrophe \cite{Mitch1984}.
In living cells, there are various microtubule 
associated proteins (MAPs) 
that either stabilize or destabilize microtubules 
and regulate microtubule dynamics both spatially and temporally
\cite{Nogales2000}. Recently, the importance of 
 MAPs in connection  with the plus end of growing MTs 
has been recognized \cite{Akhm2008}.
Stabilizing MAPs  bind to 
assembled MTs, thereby reducing the catastrophe rate or increasing 
the rescue rate. Destabilizing MAPs such as OP18/stathmin bind 
to GTP-tubulin dimers, thus decreasing the available GTP-tubulin 
concentration, which in turn decreases the growth velocity 
of the GTP-cap and makes catastrophes more likely. 
Therefore, such mechanisms can regulate basic parameters 
in our model, such as the available GTP-tubulin concentration
or the rescue rate, and we will systematically study their influence on 
the generated polymerization force for the 
three confinement scenarios (i)--(iii).

The paper is structured as follows: In Sec.\ \ref{sec:model}, the MT model and
the basic notation are introduced. We also discuss the catastrophe model and
the underlying hydrolysis mechanism in the absence and in the presence of a
resisting force. Section\ \ref{sec:simulation} deals with the simulation model. 
In Secs.\ \ref{sec:conf}, \ref{sec:const}, and
\ref{sec:elast}, results for the three scenarios (i)--(iii) 
are presented and
discussed. In Sec.\ \ref{sec:exp_wcat}, the  elastic 
obstacle is reconsidered using  an alternative 
catastrophe model based on experimental  measurements
to show  that our results are robust with 
respect to this change in the catastrophe model.
In Sec.\ \ref{sec:force-velocity-theta},  
we show that our results are also robust with respect to 
possible  generalization of the 
 force-velocity relation by introducing load-distribution factors. 
Section\ \ref{sec:discuss} contains a final discussion
and  outlook.

\section{Microtubule model}\label{sec:model}

\subsection{Single MT dynamics} 

The MT dynamics in the presence of its dynamic
instability is described in terms of probability densities and switching
rates \cite{Verd1992,Dogt1993PRL}. 
In the growing  state, a MT polymerizes with average velocity $v_{+}$. The MT
stochastically switches from a state of growth ($+$) 
to a state of shrinkage ($-$) with the
catastrophe rate $\omega_c$. In the 
shrinking state,  it rapidly depolymerizes
with an average velocity $v_{-}\simeq 3\cdot 10^{-7}\,{\rm m/s}$ (Table\
\ref{tab:parameters}). With  the rescue rate $\omega_{r}$ 
the MT stochastically switches from a state of shrinkage 
back to a state of growth. 
We model catastrophes and rescues as Poisson processes such that 
 $\langle \tau_{+} \rangle= 1/\omega_c$ and $\langle \tau_{-} \rangle
= 1/\omega_r$ are 
the average times spent in the growing  and shrinking states, 
respectively. 
The stochastic time evolution of an ensemble of independent MTs, 
growing along the
x-axis, can be  described by two coupled master equations
for the probabilities $p_{+}(x,t)$ and $p_{-}(x,t)$ of 
finding a MT with length $x$
at time $t$ in a growing or shrinking state,
\begin{align}
\partial_{t}p_{+}(x,t) &=-\omega_{c}p_{+}(x,t) + \omega_{r}p_{-}(x,t) 
    - v_{+}\partial_{x}p_{+}(x,t)
 \label{eq:p+}\\
\partial_{t}p_{-}(x,t) &=\omega_{c}p_{+}(x,t) - \omega_{r}p_{-}(x,t) 
      + v_{-}\partial_{x}p_{-}(x,t).
\label{eq:p-}
\end{align}
In the following, we will always use a 
 reflecting boundary at $x=0$: A MT shrinking back to zero length
undergoes a forced rescue instantaneously.
This corresponds to 
\begin{equation}
 v_+ p_{+}(0,t) = v_- p_{-}(0,t).
\label{eq:reflecting}
\end{equation} 
A more refined model including a nucleating state has been 
considered in Ref.\ \cite{Mulder2012}.
For a constant and fixed catastrophe rate $\omega_c$, 
eqs.\ (\ref{eq:p+}) and (\ref{eq:p-}) together 
with the boundary condition (\ref{eq:reflecting}) 
 can be solved analytically 
on the half-space $x>0$, and we can  determine  the 
overall probability density function (OPDF) of finding a MT with length $x$
at time $t$,
$P(x,t)\equiv p_{+}(x,t)+p_{-}(x,t)$ \cite{Verd1992,Dogt1993PRL}. 
The solution 
exhibits two different growth phases: 
a phase of bounded growth and a phase of
unbounded growth.

In the phase of bounded growth the average length loss
during a period of shrinkage, $v_-\langle \tau_{-} \rangle = v_-/\omega_r$,
  exceeds the average length gain during a period
of growth, $v_+\langle \tau_{+} \rangle = v_+/\omega_c$. 
The steady-state solution of $P(x,t)$ 
assumes a simple exponential form $P(x) = |\lambda|^{-1} e^{-x/|\lambda|}$ with an 
average length $\langle x \rangle= |\lambda|$
and a characteristic length parameter 
\begin{equation}
  \lambda \equiv \frac{v_{+}v_{-}}{v_{+}\omega_{r}-v_{-}\omega_{c}},
\label{eq:lambda}
\end{equation}
with $\lambda^{-1}<0$  for bounded growth \cite{Verd1992}.
The transition to the regime of unbounded growth takes place
at  $\lambda^{-1}=0$, where 
the average length gain during growth equals exactly the 
average length loss during shrinkage, 
\begin{equation}
\label{eq:condition}
  v_{+}\omega_{r}=v_{-}\omega_{c},
\end{equation} 
such that  $\langle x \rangle$ diverges.

In the regime of unbounded growth ($\lambda>0$), the
average length gain during a period of growth is larger than the average
length loss during a period of shrinkage. There  is no steady state solution,
and for long times $P(x,t)$ asymptotically approaches a Gaussian
distribution \cite{Verd1992}
\begin{equation}
   P(x,t) \approx \frac{1}{2\sqrt{\pi D_Jt}}
      \exp\left(-\frac{(x-Jt)^2}{4D_Jt} \right)
 \label{Gauss}
\end{equation}
 centered on  an
 average length which approaches linear 
 growth $\langle x \rangle \approx Jt$ with 
a mean velocity $J$ and with diffusively growing width
$\langle x^2 \rangle-\langle x \rangle^2 \approx 2D_Jt$  
with a diffusion constant $D_J$. 
The average growth velocity is given by 
\begin{equation}
J = \frac{v_{+}\omega_{r} - v_{-}\omega_{c}}{\omega_{r}+\omega_{c}}
\label{J}
\end{equation}
because the asymptotic probabilities to be in a growing or shrinking 
state are $\pi_+ = \omega_r/(\omega_c+\omega_r)$  and  
$\pi_- = \omega_c/(\omega_c+\omega_r)$, respectively. 
The diffusion constant $D_J$ is 
\begin{equation}
    D_{J}=\frac{\omega_c\omega_r(v_++v_-)^2}{(\omega_c+\omega_r)^3}.
\label{DJ}
\end{equation}

 The transition between the two
growth phases can be achieved by changing one of the four parameters of
MT growth, $\omega_c$, $\omega_r$, $v_+$, or $v_-$.  
In the following, we will use catastrophe models, where the 
catastrophe rate  $\omega_{c}$ is a function of the growth velocity 
$v_{+}$,
which in turn is determined by 
the GTP-tubulin concentration via the GTP-tubulin on-rate 
$\omega_{\text{on}}$ (assuming a 
fixed off-rate $\omega_{\text{off}}$). 
Moreover, experimental data suggest that  $v_{-}$ is
fixed to values close to $\sim 10^{-7}$ m/s (Table\ \ref{tab:parameters}).
As a consequence,
 there are two  tunable control  parameters left,  
the GTP-tubulin concentration or, equivalently, the tubulin 
on-rate $\omega_{\text{on}}$ and the
rescue rate $\omega_r$. These are the control parameters  
we will explore for MTs in confinements and under force. 
These parameters are also targets for regulation by MAPs, 
such as OP18/stathmin, which reduces $\omega_{\text{on}}$ by 
binding to GTP-tubulin dimers or MAP4, which increases 
the rescue rate $\omega_r$.

\subsection{Force-dependent catastrophe rate}

In a growing state, GTP-tubulin dimers are attached to any of the 13
protofilaments with the rate $\omega_{\text{on}}$, which is directly related
to the GTP concentration. We explore 
a regime $\omega_{\text{on}} = 30,...,100\,{\rm s^{-1}}$, see 
Table\ \ref{tab:parameters}. GTP-tubulin dimers are detached with the
rate $\omega_{\text{off}}=6\,{\rm s}^{-1}$ \cite{Janson2004} such that 
 we can typically assume $\omega_{\text{on}} \gg \omega_{\text{off}}$. 
In the absence of force or
restricting boundaries, the velocity of growth is given by
\begin{equation}\label{eq:vel_zero}
 v_{+}(0)=d\left(\omega_{\text{on}} - \omega_{\text{off}}\right).
\end{equation}
Here $d$ denotes the effective dimer size 
$d\approx 8\text{nm}/13\approx 0.6\,{\rm nm}$.

The classical view of the MT catastrophe mechanism is based on a purely 
chemical model of catastrophes, where the  
catastrophe rate $\omega_{c}$ is determined by the hydrolysis
dynamics of GTP-tubulin \cite{Mitch1984}.
 When GTP-tubulin is incorporated
into the tip of a growing MT, it forms a stabilizing GTP-cap. In a 
chemical model, the loss of this
GTP-cap due to hydrolysis of GTP-tubulin to GDP-tubulin directly  causes a
catastrophe. However, recent
research indicates that the ``structural plasticity'' of the 
MT lattice can play  a role for the kinetics of catastrophes \cite{Kueh2009}.
This structural plasticity  mechanism 
is based on the assumption that GDP-tubulin 
 prefers a  curved configuration, which generates additional mechanical 
stresses in the MT by hydrolysis. 
Also in the presence of structural plasticity, the loss of the 
GTP-cap has a destabilizing effect, but the kinetics leading to 
a catastrophe can be more complicated because 
the initiation of a catastrophe event is similar to the nucleation of a
crack in the stressed MT lattice within this model. 
 In this article, we focus on purely chemical
catastrophe models and neglect mechanical effects on the 
catastrophe kinetics.

Within a chemical catastrophe model, the loss of the
GTP-cap due to hydrolysis of GTP-tubulin to GDP-tubulin
triggers a catastrophe immediately. 
Therefore, the catastrophe rate $\omega_{c}$ is given by the
first-passage rate to a state with vanishing GTP-cap and has been discussed
within a model with cooperative hydrolysis \cite{Flyv1994PRL,Flyv1996PRE},
where GTP-tubulin is hydrolyzed by a combination of random and vectorial
mechanisms;
similar models have also been discussed for hydrolysis in F-actin 
\cite{Kier2009,Kier2010}.
In random hydrolysis, GTP-tubulin is hydrolyzed at a random site
within the GTP-cap with a rate per length $r\backsimeq 3.7 \cdot 10^{6}\, {\rm
  m^{-1}s^{-1}}$, while in vectorial hydrolysis, only GTP-tubulin with
adjacent GDP-tubulin is hydrolyzed. 
This results in hydrolysis fronts
propagating through the microtubule with average velocity $v_{h}\backsimeq 4.2
\cdot 10^{-9}\,{\rm m/s}$. The inverse catastrophe rate can then be calculated
as the mean first-passage time to a state with zero cap length, as a function
of hydrolysis parameters and $v_{+}$. With $v=v_{+}-v_{h}$,
$D=0.5d(v_{+}+v_{h})$ and $\gamma=0.5vD^{1/3}r^{-1/3}$ the exact analytical
result for the dimensionless catastrophe rate
$\alpha=\omega_{c}D^{-1/3}r^{-2/3}$ is given by the smallest solution of
\begin{equation}
 \label{eq:wcat}
 Ai'(\gamma ^{2}-\alpha) =- \gamma Ai(\gamma^{2} -\alpha).
\end{equation}
Here $Ai$ denotes the first Airy function and $Ai'$ its derivative
\cite{Abramowitz}. We solved eq.\ (\ref{eq:wcat}) numerically and obtained a
high order polynomial for the function $\alpha=\alpha(\gamma)$. 
This polynomial is used in simulations
and analytical calculations to compute the catastrophe rate  
$\omega_{c}=\alpha(\gamma)D^{1/3}r^{2/3}$ 
as a function of the growth velocity $v_{+}$,
while the hydrolysis parameters $v_h$ and $r$ are fixed.

Under a force $F$, the tubulin on-rate 
$\omega_{\text{on}}$ is modified by an additional Boltzmann
factor \cite{Pes1993} and the force-dependent growth velocity  becomes
\begin{equation}
\label{eq:vel_F}
 v_{+}(F)=d\left[\omega_{\text{on}}\exp{(-Fd/k_{B}T}) - \omega_{\text{off}}\right].
\end{equation}
Here $Fd$ is the work that has to be done against the force $F$ to incorporate
a single dimer of size $d$; $k_{B}$ is the Boltzmann constant and $T=300$ K the
temperature. In the following we use the dimensionless force 
\begin{equation}
   f \equiv F/F_{0}~~\mbox{with}~~F_{0}=k_{B}T/d,
\label{f}
\end{equation}
in terms of which  the force-dependent growth velocity is given by 
\begin{equation}
   v_+(f) = d\left[ \omega_{\text{on}}e^{-f} - \omega_{\text{off}}\right].
\label{eq:vel_f}
\end{equation}
The  characteristic force $F_0$ has a value 
$F_{0}=k_{B}T/d\approx 7\, {\rm pN}$.  
The dimensionless stall force 
\begin{equation}
 f_{\text{stall}}=\ln\left(\omega_{\text{on}}/\omega_{\text{off}}
\right)
\label{eq:fstall}
\end{equation}
 is  defined by the 
condition of vanishing growth velocity $v_+(f_{\text{stall}})=0$.
We typically have $f_{\text{stall}}\simeq 1.5,...,3$ or $F_{\text{stall}}\simeq
10,...,20\,{\rm pN}$ for $\omega_{\text{on}} =
30,...,100\,{\rm s^{-1}}$.
The stall force is the maximal force that the MT can generate in 
the {\em absence} of catastrophes. 
We will investigate how the forces that can be generated in the 
presence of catastrophes compare to this stall force.

The velocity-dependence of the 
 catastrophe rate as calculated from eq.\ (\ref{eq:wcat})
gives rise to a force-dependence
 $\omega_{c}=\omega_c(v_+(f))$.
We assume that this is the only effect of force on the catastrophe rate
\cite{Janson2003}. 
As a result, the catastrophe rate  increases exponentially, when
$v_{+}(f)$ is decreased by applying a  force $f$, but a finite value is maintained at
$v_{+}(f)=0$, which is $\omega_{c}(v_{+}=0)\approx 2.9\,{\rm s}^{-1}$. We
assume that $v_{-}$ is independent of force.
For qualitative approximations, 
the force-dependence of the catastrophe rate can be 
described by an exponential increase above the characteristic force $F_0$,
\begin{equation}
   \omega_{c}(f) \sim \omega_{c}(f\!=\!0)e^{f}, 
\label{eq:wcat_app}
\end{equation}
In Sec.\  \ref{sec:exp_wcat} we introduce an alternative catastrophe 
model which is based  on experimental  measurements.
The exponential approximation (\ref{eq:wcat_app})
  applies to the catastrophe model described above as well as to 
the  alternative catastrophe model,
see Fig.\ \ref{fig:wcat_compare}.
Our results are robust for all catastrophe models with an 
exponential increase above  the characteristic force $F_0$. 
Our results do not directly apply to more elaborate 
 multi-step catastrophe models with more than two MT states
 \cite{Gardner2011}.

\section{Simulation Model}\label{sec:simulation}

In the simulations we  solve the stochastic Langevin-like 
 equations of motion for the length
$x(t)$ of a single MT using numerical integration with fixed time steps 
$\Delta t$
and including stochastic  switching between growth and shrinkage. In a
growing state $x(t)$ is increased by $v_{+}\Delta t$, while in a state of
shrinkage, it is decreased by $v_{-}\Delta t$. In the growing state,
  $v_{+}$ is
calculated from eq.\ (\ref{eq:vel_zero}) for zero force and from eq.\
(\ref{eq:vel_F}) under force. In each time step a uniformly distributed
random number $\xi \in [0,1]$ is compared to $\omega_{r,c}\Delta t$. If
$\xi<\omega_{r,c}\Delta t$ the MT changes its state of growth.  The
catastrophe rate $\omega_{c}$ is calculated from the high order polynomial
obtained from eq.\ (\ref{eq:wcat}) as mentioned above. To assure
$\omega_{r,c}\Delta t\leq 1$ we used a time step $\Delta
t=0.1\,{\rm s}$. 
During the simulations all parameters of growth, $d=0.6\,{\rm nm}$,
$r=3.6\cdot 10^{6}\,{\rm m^{-1}s^{-1}}$, $v_{h}=4.2\cdot 10^{-9}\,{\rm m/s}$,
$k_{B}=1.38\cdot 10^{-23} \,{\rm J/K}$, $T=300\, {\rm K}$,
  and $\omega_{\text{off}}=6\,{\rm s}^{-1}$,
are fixed, see Table \ref{tab:fixed_parameters}, 
except for $\omega_{\text{on}}$, which is varied in 
the range $\omega_{\text{on}} = 30-100\,{\rm s^{-1}}$, and 
$\omega_r$, which is varied in a range $\omega_{r} = 0.03-0.2\,{\rm
  s^{-1}}$, see Table \ref{tab:parameters}. 
Averages are taken over many realizations of stochastic 
trajectories.

\section{Confinement between fixed rigid walls}\label{sec:conf}

\begin{figure}
\includegraphics[width=\columnwidth ]{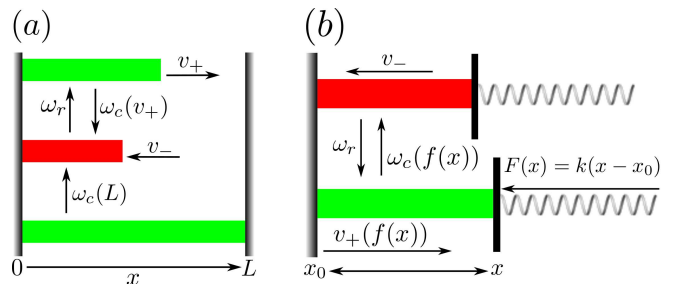}
\caption{ (a): Schematic representation of the confinement and possible MT
  configurations. From top to bottom: A MT growing with $v_{+}$; MT shrinks
  with $v_{-}$. MT in a state of growth and stuck to the boundary wall with
  $v_{+}=0$ and $\omega_{c,L}$.  (b): Schematic representation of a single MT
  growing against the elastic obstacle. From top to bottom: MT shrinks with
  $v_{-}$. MT under force $F(x)=k\left(x-x_{0}\right)$ with $f(x)\equiv
  F(x)/F_{0}$, $v_{+}[f(x)]$, and force-dependent catastrophe rate
  $\omega_{c}[f(x)]$. \label{fig:model}}
\end{figure}

A single MT is confined to a one-dimensional box of  fixed length
$L$ with rigid boundary walls at $x=0$ and $x=L$  as 
shown schematically in Fig.\ \ref{fig:model}(a)
\cite{Holy1994,Dogt1998PRL}. There is no force acting on the MT
but  within  the box catastrophes are induced upon hitting 
the rigid walls.
We propose the following mechanism for these wall-induced 
catastrophes:
 When the MT
hits the boundary at $x=L$, its growth velocity $v_{+}$ has to 
reduce  to zero, which leads to an increase of the  catastrophe
rate to $\omega_{c,L}\equiv\omega_{c}(v_{+}=0)$. Since $\omega_{c,L}$ is
finite,  wall-induced catastrophes are not instantaneous
 but the MT sticks for an average 
 time $1/\omega_{c,L}$ to the boundary before the catastrophe, 
which is in contrast to previous studies
\cite{Gov1993}. For the average time spent at the boundary before 
a catastrophe, we find
$\omega_{c,L}^{-1}\approx 0.29\,{\rm s}$.
The catastrophe rate at the wall, $\omega_{c,L}$,
 is much higher than the bulk 
catastrophe rate $\omega_c(v_+)$. 
For $\omega_{\text{on}} = 50\,{\rm s^{-1}}$ we find 
$\omega_{c,L}/\omega_{c} \simeq 2300$.

To include the mechanism of wall-induced catastrophes into 
the description by master equations, we introduce
 the probabilities $Q_{+}$ and 
$Q_{-}$ of finding the MT stuck to the boundary in a
growing state and in a shrinking state, respectively. The stochastic time
evolution of $Q_{+}(t)$ and $Q_{-}(t)$ is given by:
\begin{align}
  \partial_{t}Q_{+}(t) &=-\omega_{c,L} Q_{+}(t) + \omega_{r}Q_{-}(t) 
   + v_{+}p_{+}(L) \label{eq:q+}\\
 \partial_{t}Q_{-}(t) &=+\omega_{c,L} Q_{+}(t) - \omega_{r}Q_{-}(t) 
  - \frac{v_{-}}{\Delta} Q_{-}(t) \label{eq:q-}.
\end{align}
The quantity $v_{+}p_{+}(L)$ is the flow of probability from the interior of the
confining box  onto its boundary and is given by the solution of eq.\
(\ref{eq:p+}) and (\ref{eq:p-}) for $x=L$, while $(v_{-}/\Delta) Q_{-}$ is
the probability current from the boundary back into the interior, where
$\Delta$ denotes a small interval in which the flow $v_{-}Q_{-}$ can be
measured.
This implies that there is a boundary condition 
$v_- p_{-}(L,t) = (v_{-}/\Delta) Q_{-}$ for the backward current density 
at $x=L$, in addition to the reflecting boundary condition 
(\ref{eq:reflecting}) at $x=0$. 
An identical model for wall-induced catastrophes 
has been introduced  in Ref.\ \cite{Mulder2012} recently.

 In the steady state and in the limit $\Delta \approx 0$ we
find
\begin{align}
Q_{+} &\approx \frac{v_{+}}{\omega_{c,L}}p_{+}(L) 
    \label{eq:qs+} \\
Q_{-} &\approx 0, \label{eq:qs-}
\end{align}
and $v_- p_{-}(L,t) = (v_{-}/\Delta) Q_{-} = v_+ p_{+}(L)$. 
Eq.\ (\ref{eq:qs+}) shows that there is a non-zero probability $Q_{+}$ 
of finding a MT in a state of growth and stuck to the boundary, which is
given by the flow of probability from the interior of the confining box 
 onto its
boundary divided by the average time being stuck to the boundary. In contrast,
eq.\ (\ref{eq:qs-}) states that there is no MT  in a shrinking state 
and stuck to the wall. This is intuitively clear since a MT undergoing a
catastrophe begins to shrink instantaneously. 
In the steady state, we solve eqs.\ (\ref{eq:p+}),
(\ref{eq:p-}) and (\ref{eq:qs+})  simultaneously with the 
additional normalization 
$\int^{L}_{0}(p_{+}(x)+p_{-}(x))dx + Q_{+}=1$. We find
$v_+p_+(x) = v_-p_-(x)$ and 
\begin{align}
P(x)&=  N e^{x/\lambda}  \left(1 + \frac{v_{+}}{v_{-}}\right)
  \label{eq:conf_p} \\
Q_{+} &= N \frac{v_{+}}{\omega_{c,L}} e^{L/\lambda}
\label{eq:conf_q}
\end{align}
with $\lambda$ from eq.\ (\ref{eq:lambda}) and 
a normalization
\begin{equation}
  N^{-1}=\lambda\left( 1+\frac{v_{+}}{v_{-}} \right)
      \left(e^{L/\lambda}-1 \right)
     + \frac{v_{+}}{\omega_{c,L}}e^{L/\lambda}.
\label{eq:conf_norm}
\end{equation}
Equation (\ref{eq:conf_p}) shows that we find an exponential
OPDF $P(x)$ in confinement with the same characteristic length 
$|\lambda|$. If the growth is unbounded in the absence
of confinement, which corresponds to $\lambda^{-1}>0$, the OPDF 
is exponentially increasing;
if the growth is bounded in the absence of confinement, which corresponds 
to $\lambda^{-1}<0$, the OPDF remains exponentially 
decreasing in confinement.
The same result has been obtained in Ref.\ \cite{Gov1993} 
within a discrete growth model.
In independent {\em in vivo} experiments,  both 
exponentially increasing \cite{Komarova2002} and exponentially decreasing
OPDFs \cite{Verd1992} have been found.

In the following we focus on the case $\lambda^{-1}>0$ of exponentially 
increasing OPDFs.
In the steady state, 
the average length of a MT within the confining box
 is  given by 
\begin{align}
\langle x \rangle& =\int^{L}_{0}xP(x)dx + Q_+L    \nonumber\\
    & = N \left\{ \left( 1 +\frac{v_{+}}{v_{-}}\right ) 
   \lambda^2 \left[1+   e^{L/\lambda} \left(\frac{L}{\lambda}-1\right) \right]
        \right.
\nonumber\\
  &~~~~~\left.
     + L \frac{v_{+}}{\omega_{c,L}}e^{L/\lambda} \right\}.
\label{eq:meanlength}
\end{align}
In the limit of instantaneous 
wall-induced catastrophes, $Q_+ \approx 0$, 
we obtain 
\begin{equation}
 \frac{\langle x \rangle}{L} 
   \approx  \frac{1}{1-e^{-L/\lambda}}-\frac{\lambda}{L},
\label{eq:meanlength2}
\end{equation}
 i.e., the average MT length 
$\langle x \rangle/L$ depends on the two control  parameters 
$\omega_r$ and  $\omega_{\text{on}}$ only via  the ratio $L/\lambda$. 
This scaling property is lost if wall-induced catastrophes are not 
instantaneous
because eq.\ (\ref{eq:meanlength}) then 
exhibits additional $v_+$- and thus $\omega_{\text{on}}$-dependencies.
Within our model the increased catastrophe rate 
at the boundary gives rise  to an increased 
overall average catastrophe rate
\begin{equation}
  \omega_{\text{c,eff}} = \omega_c(v_+)+ Q_+( \omega_{c,L}-\omega_c(v_+)),
\end{equation} 
for which we find 
$\omega_{\text{c,eff}} \simeq 0.03\,{\rm s^{-1}}$ for $L=1\,{\rm \mu m}$ and 
$\omega_{\text{c,eff}} \simeq 0.006\,{\rm s^{-1}}$ for $L=10\,{\rm \mu m}$
as compared to  $\omega_{c}\simeq 0.0015\,{\rm s^{-1}}$ for these 
conditions.

\begin{figure}
   \includegraphics[width=\columnwidth]{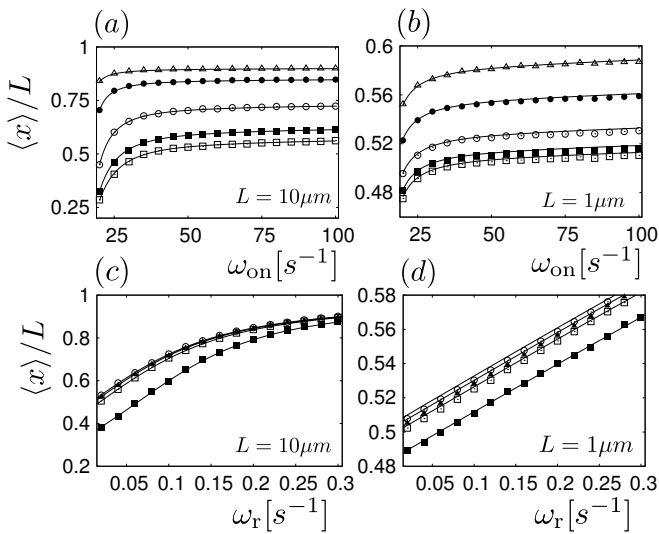}
  \caption{ The average length $\langle x\rangle$ as a function of
    $\omega_{\text{on}}$ and $\omega_{r}$ for confinement by fixed rigid walls. 
        Data points are results from 
    stochastic simulations, lines are analytical results (\ref{eq:meanlength}). 
    Top row: The average length $\langle x\rangle$ as a function of
    $\omega_{\text{on}}$ for different values of $\omega_{r}=0.03\,{\rm
      s}^{-1}(\boxdot), 0.05\,{\rm s}^{-1}(\blacksquare), 0.1\,{\rm
      s}^{-1}(\odot), 0.2\,{\rm s}^{-1}(\bullet)$ and $0.3\,{\rm
      s}^{-1}(\triangle)$. (a) $L=10\,{\rm \mu m}$.  
     (b) $L=1\,{\rm \mu m}$.    Lower row: The
    average length $\langle x\rangle$ as a function of $\omega_{r}$ for
    different values of $\omega_{\text{on}}
    =25\,{\rm s}^{-1}(\blacksquare), 50\,{\rm  s}^{-1}(\boxdot),
        75\,{\rm s}^{-1}(\blacktriangle), 100 \,{\rm
      s}^{-1}(\boxdot)$.  (c) $L=10\,{\rm \mu m}$. 
     (d) $L=1\,{\rm \mu m}$.\label{fig:conf_x}}
\end{figure}

\begin{figure}
   \includegraphics[width=\columnwidth]{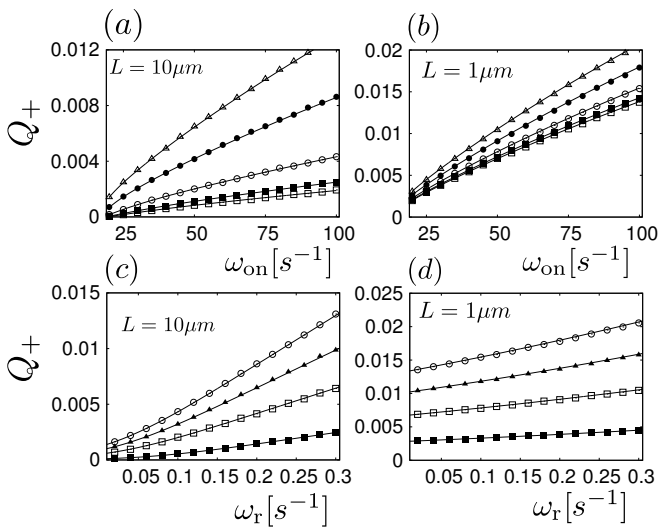}
  \caption{  The probability $Q_+$ to find the MT at the wall 
  as a function of
    $\omega_{\text{on}}$ and $\omega_{r}$ for confinement by fixed rigid walls. 
      Data points are results from 
    stochastic simulations, lines are analytical results (\ref{eq:qs+}). 
   Top row: $Q_{+}$ as a function of $\omega_{\text{on}}$ for
    different values of $\omega_{r}=0.03\,{\rm s}^{-1}(\boxdot), 0.05\,{\rm
      s}^{-1}(\blacksquare), 0.1\,{\rm s}^{-1}(\odot), 0.2\,{\rm
      s}^{-1}(\bullet)$ and $0.3\,{\rm s}^{-1}(\triangle)$. 
   (a) $L=10\,{\rm \mu m}$. (b) $L=1\,{\rm \mu m}$. \\
   Lower row: $Q_{+}$ as a function of
    $\omega_{r}$ for different values of $\omega_{\text{on}}=25\,{\rm
      s}^{-1}(\blacksquare), 50\,{\rm s}^{-1}(\boxdot), 75\,{\rm
      s}^{-1}(\blacktriangle), 100\,{\rm s}^{-1}(\boxdot)$.  (c)
     $L=10\,{\rm \mu m}$.  (d)  $L=1\,{\rm \mu m}$. 
\label{fig:conf_q}}
\end{figure}

We set the length of the confining box to  $L=1\,{\rm \mu m}$ 
 and $L=10\,{\rm \mu m}$, which are
typical length scales in experiments \cite{Faiv2008,Schek2007} and cellular
environments \cite{daga2006}, and we calculate $\langle x\rangle $ and $Q_{+}$ as
functions of $\omega_{\text{on}}$ and $\omega_{r}$. 
The parameter regimes displayed 
in Figs.\ \ref{fig:conf_x} and \ref{fig:conf_q}
correspond to regimes $L/\lambda \gg 1$ for $L=10\,{\rm \mu m}$
and $L/\lambda \ll 1$ for $L=1\,{\rm \mu m}$.
Results obtained from stochastic 
simulations agree  with analytical findings (Figs.\ \ref{fig:conf_x} and
\ref{fig:conf_q}). It is clearly visible that the size $L$ of the confinement
has a significant influence on $\langle x\rangle$, mainly 
via the ratio $L/\lambda$. 

The probability 
 $Q_{+}$ to find the MT at the wall 
increases with increasing rates in the range of $Q_{+}\approx
0,...,0.03$ and exhibits only a weak dependency on $L$, see 
Figs.\ \ref{fig:conf_q}. Even for maximum rates,
the probability of finding a MT in a growing state and stuck to the wall is
limited to several percent, due to the large catastrophe rate $\omega_{c,L}$ 
at $x=L$.
Therefore, in most cases  wall-induced catastrophes can be viewed 
as instantaneous, and
 the approximation (\ref{eq:meanlength2}) works well. 
For increasing on-rate  $\omega_{\text{on}}$
or  rescue rate $\omega_{r}$, the ratio $L/\lambda$ approaches 
$L/\lambda \approx L\omega_r/v_-$ from below.
According to the approximation 
  (\ref{eq:meanlength2}), the mean length 
$\langle x\rangle$ then increases and approaches
$\langle x\rangle/L \approx  {1}/{(1-e^{-L\omega_r/v_-})}-{v_-}/{L\omega_r}$
from below.
For $L=10\,{\rm \mu m}$, we have $L/\lambda \gg 1$ and 
the length distribution is exponential, 
$P(x) \sim e^{x/\lambda}$. The ratio 
$\langle x\rangle/L$ saturates at a high value $\langle x\rangle/L 
  \approx  0.7,...,0.9$ (Figs.\ \ref{fig:conf_x} (a),(c)).
For $L/\lambda \gg 1$ the MT length distribution becomes very 
narrow around the maximal length $L$.
 In contrast, for  $L=1\,{\rm \mu m}$, we have $L/\lambda \ll 1$, 
and  $L$ is too small to
establish the characteristic exponential decay of the 
 length distribution.
The length distribution $P(x)$ is almost uniform, and the ratio 
 $\langle x\rangle/L \approx 0.5,...,0.6$ deviates only slightly
from the result $\langle x\rangle/L = 1/2$ characteristic 
for a broad uniform distribution 
(Figs.\ \ref{fig:conf_x}(b),(d)).

\section{Constant force}\label{sec:const}

In the second scenario a constant force $F$ is applied to the MT 
and the right boundary is removed, so that the MT is allowed to grow 
on $x\in [0,\infty[$. According to eq.\ (\ref{eq:vel_f})
 the  growth velocity under
force is smaller, but it remains constant for
fixed $f$. With eq.\ (\ref{eq:wcat}) this results in a higher, but also
constant, catastrophe rate $\omega_{c}[v_{+}(f)]>\omega_{c}[v_{+}(0)]$. Since
$v_{-}$ and $\omega_{r}$ are independent of force, the stochastic dynamics of
the MT is described by eq.\ (\ref{eq:p+}) and (\ref{eq:p-}) with the same
solutions $P(x,t)$ as in the absence of force, but  with a decreased
velocity of growth $v_{+}(f)$ and an increased catastrophe rate
$\omega_{c}(f)$ \cite{Dogt1993PRL,Verd1992}. 
In particular, we still find two regimes, a regime of bounded 
growth and a regime of unbounded growth.

In the regime of bounded growth
$P(x,t)$ is again exponentially decreasing, and the force-dependent average
length is $\langle x(f) \rangle= |\lambda(f)|$
with the 
corresponding force-dependent length parameter 
\begin{equation}
  \lambda(f)  \equiv
  \frac{v_{+}(f)v_{-}}{v_{+}(f)\omega_{r}-v_{-}\omega_{c}(f)}
\label{eq:lambdaf}
\end{equation}
as compared to  eq.\ (\ref{eq:lambda}) in the absence of force. 
In the regime of unbounded
growth $\langle x(f) \rangle$ increases linearly in time with the 
force-dependent mean velocity $J(f) = [v_{+}(f)\omega_{r} -
v_{-}\omega_{c}(f)]/[\omega_{r}+\omega_{c}(f)]$, cf.\ eq.\ (\ref{J}). 
The MT length distribution 
$P(x,t)$ assumes 
again a Gaussian form (\ref{Gauss}) where also the diffusion constant 
$D_J(f)$ follows the same eq.\ (\ref{DJ}) with 
  force-dependent  growth velocity 
$v_{+}(f)$ and  catastrophe rate $\omega_{c}(f)$.

\begin{figure}
  \includegraphics[width=\columnwidth]{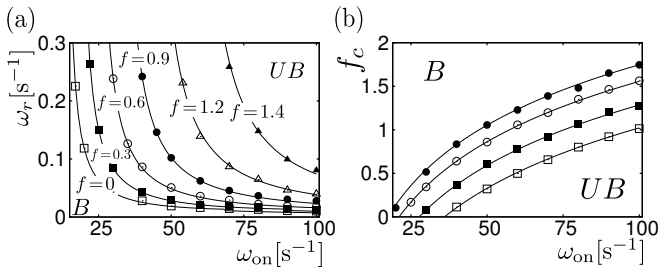}
\caption{ (a): Phase boundary between bounded 
   (B) and unbounded growth (UB) as
  a function of $\omega_{\text{on}}$ and $\omega_{r}$ 
   for MT growth under constant force. 
  Data points for $f=0(\boxdot),
  0.3(\blacksquare), 0.6(\odot), 0.9(\bullet),
  1.2(\triangle),1.4(\blacktriangle)$
    represent results from simulations, lines represent solutions of
  $v_{+}(f)\omega_{r}=v_{-}\omega_{c}(f)$ for a constant force $f$.  
   (b): 
   Critical force 
  $f_{c}$ as a function of $\omega_{\text{on}}$ for $\omega_{r}=0.03\,{\rm
    s}^{-1}(\boxdot),0.05\,{\rm s}^{-1}(\blacksquare),0.1\,{\rm
    s}^{-1}(\odot),0.2\,{\rm s}^{-1}(\bullet)$.  Data points 
    represent results from simulations, lines represent the solution of
  eq.\ (\ref{eq:condition_fc}) for a fixed combination of
  $\omega_{\text{on}}$ and $\omega_{r}$.
\label{fig:Fconst}}
\end{figure}

In the presence of a constant force $f$, the transition between 
bounded and unbounded growth is governed by the 
 force-dependent parameter $\lambda(f)$.
The regimes of bounded and
unbounded growth are now separated by the condition
$\lambda^{-1}(f)=0$, 
 which is shifted compared to the case
$f=0$, see Fig.\ \ref{fig:Fconst}(a).
The inverse length parameter $\lambda^{-1}(f)$ 
is a monotonously decreasing function
of force $f$ and changes sign from positive to negative values for 
increasing force $f$. Therefore $\lambda^{-1}(f_c)=0$ or 
\begin{equation}\label{eq:condition_fc}
 v_{+}(f_c)\omega_{r} =v_{-}\omega_{c}(f_c),
\end{equation}
defines a critical 
force for the transition from unbounded to bounded growth. 
A single MT exhibiting unbounded growth ($\lambda^{-1}(0)>0$) in the 
absence of force undergoes a transition to bounded growth with 
$\lambda^{-1}(f)<0$  by applying a 
supercritical force $f>f_c$.
 On the other hand, starting with a
combination of on-rate $\omega_{\text{on}}$ and rescue rate 
 $\omega_{r}$ and a force $f$, which results in
bounded growth with $\lambda^{-1}(f)<0$, 
the MT can still enter the regime of
unbounded growth by increasing 
 $\omega_{\text{on}}$ or $\omega_{r}$ so that
the force $f$ becomes subcritical,   $\lambda^{-1}(f)>0$ or $f<f_c$.

Rewriting condition (\ref{eq:condition_fc}) as $v_+(f_c) = v_-
\omega_c(f_c)/\omega_r >0$ and using that $v_+(f)$ decreases 
with $f$, it follows that  the critical 
force is always smaller than the stall force, $f_c < f_{\text{stall}}$, which 
satisfies $v_+(f_{\text{stall}}) =0$, and it approaches the stall force 
only for vanishing catastrophe rate.  
Qualitatively, we can obtain an explicit result for  the critical force $f_c$
by
using the approximations of an exponentially decreasing  growth velocity,
 $v_{+}(f)\approx
v_{+}(0)e^{-f}$, which is valid for $\omega_{\text{on}}\gg \omega_{\text{off}}$
(see eq.\ (\ref{eq:vel_f})), and  an exponentially increasing 
catastrophe rate above the  characteristic force $F_0$, 
eq.\ (\ref{eq:wcat_app}),
in the condition (\ref{eq:condition_fc}) for the critical force.
 This leads to 
\begin{equation}
  \label{eq:fc}
    f_c \sim
    \frac{1}{2} \ln\left( \frac{v_+(0) \omega_r}{v_- \omega_c(0)}\right)
    \sim 
  \frac{1}{2} \ln\left( \frac{\omega_{\text{on}}d \omega_r}
    {v_- \omega_c(0)}\right)
\end{equation}
which shows that the critical force grows approximately logarithmically 
with on-rate $\omega_{\text{on}}$  (note that the catastrophe rate 
in the absence of force decreases with $\omega_{\text{on}}$ as 
$\omega_c(0)\propto 1/\omega_{\text{on}}$ \cite{Flyv1996PRE})
and rescue rate $\omega_{r}$.
A negative $f_c$  for  small on-rates  and rescue rates signals 
that the MT is for all forces $f>0$ in the bound phase. 
In Fig.\ \ref{fig:Fconst}(b) we show  exact results 
for the critical force $f_{c}$  as a function of the 
on-rate $\omega_{\text{on}}$ and for different rescue rates
$\omega_{r}$
from solving condition (\ref{eq:fc}) numerically 
 and from stochastic  simulations.
 Agreement between both methods  is good.

The condition $\lambda^{-1}(f)=0$ specifies the boundary between 
bounded and unbounded growth at a given force $f$. 
In Fig.\ \ref{fig:Fconst}(a), 
the resulting phase boundary is shown as a function of  $\omega_{\text{on}}$ and
$\omega_{r}$.  There is good agreement between
numerical solutions of $\lambda^{-1}(f)=0$  and 
stochastic simulations. With
increasing force, the boundary between the two regimes of growth shifts to
higher values of $\omega_{\text{on}}$ and $\omega_{r}$,
 and forces up to $F\sim 1.4\cdot F_{0}$ can be overcome by a single MT
in the parameter regimes of $\omega_{\text{on}}$ and $\omega_{r}$ considered.

\section{Elastic force}\label{sec:elast}
\label{sec:elastic}

In the third scenario, an elastically coupled barrier 
is placed in front of the
MT as shown in  Fig.\ \ref{fig:model}(b), 
which  models the optical traps used in Refs.\
\cite{Schek2007, Laan2008}  or 
the elastic cell cortex {\em in vivo}. 
If the barrier is displaced from its
equilibrium position $x_{0}$ by the growing MT with length $x>x_{0}$, it
causes a force $F(x)=k(x-x_{0})$ resisting further growth. For $x<x_{0}$ there
is no force.
We use $x_0= 0\,{\rm \mu m}$ in the case of vanishing rescue rate and
$x_0= 10\,{\rm \mu m}$ in the case of finite rescue rate
and a  spring constant $k$ in the range 
$10^{-7}\,{\rm N/m}$ (soft) to $10^{-5}\,{\rm N/m}$ 
(stiff as in the optical trap 
experiments in \cite{Laan2008}).

An elastic force $F(x)=k(x-x_{0})$ represents the simplest 
and most generic $x$-dependent force. Whereas for a 
confinement of fixed length or a constant force, the MT length 
$x$ was the only stochastic variable,  the force $F(x)$ itself 
is now coupled to $x$ and becomes stochastic as well. 
Therefore, not only are the MT length distributions of interest 
but also  the maximal and average polymerization forces which are generated 
during MT growth.

\subsection{Vanishing rescue rate}\label{sec:elast_wres0}

We first discuss growth in the absence of rescue events, 
$\omega_{r}=0$. This situation corresponds to optical
trap experiments \cite{Schek2007,Laan2008}, which 
 are performed on short time scales and no
rescue events are observed.  
In a state of growth the MT
grows against the elastic obstacle with velocity $v_{+}[f(x)]$ and $f(x)$
increases.  For simplicity we suppress the $x$-dependency in the notation
in the following.
At a maximal polymerization force $f_{\text{max}}$, the MT undergoes a
catastrophe and starts to shrink back to zero and the dynamics stop due to
missing rescue events. No steady state is reached. Since switching to the
state of shrinkage is a stochastic process, the maximal 
polymerization force $f_{\text{max}}$ is a stochastic
quantity which fluctuates around its average value. We calculate the average
maximal polymerization force $\langle f_{\text{max}}\rangle$ within a
mean field approach. Here $\langle ...\rangle$ denotes an ensemble average
over many realizations of the growth experiment.

Because no steady state is reached in the absence of rescue events,
we have to use a dynamical mean field approach,
which is based on the fact that the 
 MT growth velocity $dx/dt=v_{+}(f)$ 
is related to the time evolution of the force by
 $df/dt = (k/F_0)dx/dt$.  
In mean field theory, this results in the following 
 equation of motion for $\langle f\rangle$,
\begin{equation}
\frac{d}{dt}\langle f\rangle=\frac{k}{F_0} v_{+}(\langle f\rangle),
  \label{eq:time_evo0}
\end{equation}
where we used the mean field approximation 
$\langle v_{+}(f)\rangle  \approx v_{+}(\langle f\rangle)$. 
With the initial condition $\langle f\rangle(0)=0$ we find
a time evolution 
\begin{eqnarray}
 \langle f\rangle(t)&=&\ln \left[ \left( 
  1 - \omega_{\text{on}}/\omega_{\text{off}} \right) e^{-t/\tau} 
 + \omega_{\text{on}}/\omega_{\text{off}}       \right] 
\label{eq:w_r0_f(t)}\\
 &\approx& f_{\text{stall}} 
+ \ln \left[ 1 - \exp(-t/\tau)\right]
\label{eq:w_r0_f_ap}
\end{eqnarray}
with a characteristic time scale 
$\tau=F_{0}/dk\omega_{\text{off}}\approx 10^{2}...10^{4}\, {\rm s}$ for
$k\approx  10^{-5}...10^{-7}\, {\rm N/m}$.
For long times $t\gg \tau$,  eq.\ (\ref{eq:w_r0_f(t)}) approaches the 
dimensionless stall force 
$\langle f\rangle= f_{\text{stall}}$, see eq.\ (\ref{eq:fstall}), 
which is the maximal polymerization force in the 
absence of catastrophes. 
The  approximation (\ref{eq:w_r0_f_ap}) holds 
for  $\omega_{\text{on}}/\omega_{\text{off}}\gg 1$.

\begin{figure}
 \includegraphics[width=\columnwidth]{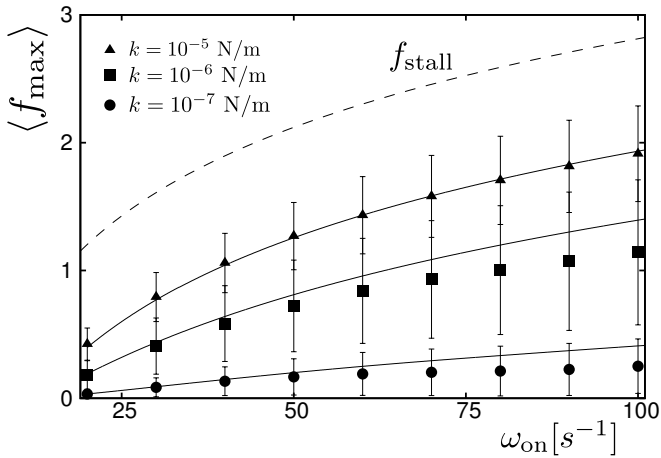}
 \caption{
   Average maximal polymerization force $\langle f_{\text{max}}\rangle$
   for an elastic obstacle and in the absence of rescues
   as a function of $\omega_{\text{on}}$ for different values of 
   $k=10^{-5}\,{\rm N/m}(\blacktriangle)$, 
  $10^{-6}\,{\rm  N/m}(\blacksquare)$ and $10^{-7}\,{\rm  N/m}
   (\bullet)$. Data points 
    represent results from simulations, solid lines
   are solutions of eq.\ (\ref{eq:w_r0_f_max}). 
 Error bars represent the  standard deviation of the 
stochastic quantity $\langle f_{\text{max}}\rangle$.
    Dashed line: dimensionless stall force
   $f_{\text{stall}}=\ln\left( \omega_{\text{on}}/\omega_{\text{off}}
   \right)$.
\label{fig:N1_FMAX}}
\end{figure}

MT growth is ended, however,  by a
catastrophe, and the average time spent in the growing state is
$t =1/\omega_{c}(\langle f_{\text{max}}\rangle)$. 
Together with eq.\
(\ref{eq:w_r0_f(t)}), this gives a 
self-consistent mean field equation for the maximal 
polymerization force $\langle f_{\text{max}}\rangle $, 
\begin{equation}
  \langle f_{\text{max}}\rangle= 
 \ln \left[ \left( 1 - \omega_{\text{on}}/\omega_{\text{off}}  \right) 
 e^{-1/\omega_{c}(\langle f_{\text{max}}\rangle)\tau} 
   +    \omega_{\text{on}}/\omega_{\text{off}}  \right]. 
\label{eq:w_r0_f_max}
\end{equation}
 The maximal polymerization force 
$\langle f_{\text{max}}\rangle$ is always smaller than the stall 
 force $f_{\text{stall}}$
as can be seen from eqs.\ (\ref{eq:w_r0_f(t)},\ref{eq:w_r0_f_ap}). Since 
$\omega_{\text{on}}/\omega_{\text{off}}\gg \omega_{c}\tau \gg 1$ 
for realistic force and parameter values, eq.\
(\ref{eq:w_r0_f_max}) can be approximated by
\begin{eqnarray}
 \langle f_{\text{max}} \rangle &\approx& \ln \left(
   \frac{\omega_{\text{on}}}{\omega_{\text{off}}\tau \omega_{c}(\langle
     f_{\text{max}} \rangle)}  \right) 
    \nonumber\\
   &=& f_{\text{stall}} 
   - \ln \left[ \tau \omega_{c}(\langle f_{\text{max}} \rangle)\right].
\end{eqnarray}
For a  catastrophe rate increasing exponentially above the
characteristic force $F_0$, eq.\ (\ref{eq:wcat_app}),
 we find
\begin{equation}
\langle f_{\text{max}} \rangle \sim  \frac{1}{2}\ln \left(
   \frac{\omega_{\text{on}}dk}{F_0 \omega_{c}(0)}\right),
\label{eq:w_r0_f_max2}
\end{equation}
i.e., the  maximal
polymerization force grows logarithmically in $\omega_{\text{on}}$
 (note that the catastrophe rate 
in the absence of force decreases  as 
$\omega_c(0)\propto 1/\omega_{\text{on}}$ \cite{Flyv1996PRE}), 
see Fig.\ \ref{fig:N1_FMAX} for $k=10^{-5}\,{\rm N/m}$. 
Within a slightly different   catastrophe
model obtained from experimental data and 
discussed in section \ref{sec:exp_wcat}, this 
logarithmic dependence  can be shown exactly.

Fig.\ \ref{fig:N1_FMAX} shows $\langle f_{\text{max}}\rangle$ as
a function of $\omega_{\text{on}}$. Analytical results from eq.\
 (\ref{eq:w_r0_f_max}) agree with numerical findings from stochastic
 simulations. The maximal polymerization force $\langle
 f_{\text{max}}\rangle$ increases with increasing $k$, see eq.\
 (\ref{eq:w_r0_f_max2}), but it remains smaller than the stall force
 $f_{\text{stall}}$. 
Stochastic simulations show considerable fluctuations of $f_{\text{max}}$, 
 which are caused by broad and
  exponentially decaying probability distributions for
  $f_{\text{max}}$ and which we quantify by measuring the standard deviation
$\langle f_{\text{max}}^2 \rangle - \langle f_{\text{max}} \rangle^2$.
 For increasing $k$, probability distributions
  become more narrow and mean field results approach the 
  simulation results  for  $\langle f_{\text{max}}\rangle$.

\subsection{Non-zero rescue rate}

For  a non-zero rescue rate 
$\omega_{r}$,  phases of
growth, in which $f(x)$ increases and which 
last $1/\omega_c(f)$ on average, are ended by catastrophes
which are followed by phases of shrinkage. Shrinking phases 
last $1/\omega_r$ on average, and during shrinkage the elastic
obstacle is relaxed and $f(x)$ decreases. After
 rescue,  the
MT switches back to a state of growth. 
In contrast to the case without rescue events, the system can 
attain a steady state. 
In this steady state, the average length
loss during shrinkage, $v_{-}/\omega_{r}$, equals the average length gain
during growth, 
$v_{+}(f)/\omega_{c}(f)$, and the MT oscillates around a time-averaged stall
length $\langle x \rangle $, which is directly related to the time-averaged
polymerization force by $\langle f \rangle = (k/F_0) (\langle x \rangle-x_0)$.
 In the following, the steady state dynamics and the
average polymerization force are characterized. We start with an 
analysis of the full master equations focusing on the stationary state 
followed by a dynamical mean field theory, which can also be applied to 
dilution experiments.

In the presence of a $x$-dependent  force $f(x)$, the master equations for the 
time evolution of   $p_{+,-}(x,t)$ become
\begin{align}
  \partial_{t}p_{+}(x,t) &=-\omega_{c}(x)p_{+}(x,t) + \omega_{r}p_{-}(x,t) 
        \nonumber\\
  &~~~~~~~~~~~~~~~~~~~ - \partial_{x}(v_{+}(x)p_{+}(x,t))
      \label{eq:pf+}\\
\partial_{t}p_{-}(x,t) &=\omega_{c}(x)p_{+}(x,t) - \omega_{r}p_{-}(x,t) 
     + v_{-}\partial_{x}p_{-}(x,t),
\label{eq:pf-}
\end{align}
which differ from  eqs.\ (\ref{eq:p+}) and (\ref{eq:p-}) by the  
$x$-dependence
of growth velocity  and catastrophe rate.
Both  growth velocity $v_+(x)=v_+[f(x)]$
and catastrophe rate $\omega_c(x)=\omega_c\{v_+[f(x)]\}$
become $x$-dependent via their force-dependence.
Therefore, also  the force-dependent length
parameter $\lambda(f)$ from  eq.\ (\ref{eq:lambdaf}) 
becomes $x$-dependent via 
its force-dependence, 
$\lambda(x) = \lambda[f(x)]$.
Eqs.\  (\ref{eq:p+}) and (\ref{eq:p-}) are  supplemented 
by reflecting boundary conditions  $v_+(0) p_{+}(0,t) = v_- p_{-}(0,t)$ 
at $x=0$, similar to eq.\ (\ref{eq:reflecting}).

For the steady state,
 eqs.\ (\ref{eq:pf+}) and  (\ref{eq:pf-})
are solved on the half-space $x>0$ with reflecting boundary conditions
at $x=0$, and 
we can calculate the 
 overall MT length distribution $P(x)= p_+(x)+p_-(x)$ explicitly,
\begin{equation}\label{eq:Pf}
  P(x)=N\left( 1+\frac{v_{-}}{v_{+}(x)}\right)
       e^{x_0/\lambda(0)}
       \exp \left[  \int^{x}_{x_0} dx'/\lambda(x')  \right]
\end{equation}
with a normalization
\begin{equation}\label{eq:norm_f}
 N^{-1}= \int _{0}^{\infty} dx\left( 1+\frac{v_{-}}{v_{+}(x)}\right)
       e^{x_0/\lambda(0)}
         e^{\int^{x}_{x_0} dx'/\lambda (x')},
\end{equation}
where $\lambda(x) = \lambda(f\!=\!0)$ in the force-free region  $x<x_0$ and 
$\lambda(x) = \lambda[f(x)]$ for $x>x_0$ and, likewise,
$v_+(x) = v_+(f\!=\!0)$ for  $x<x_0$  and $v_+(x) = v_+[f(x)]$ for 
$x>x_0$.
This implies $e^{x_0/\lambda(0)}e^{\int^{x}_{x_0} dx'/\lambda (x')} =
e^{x/\lambda(0)}$ and, thus, a simple exponential dependence 
of $P(x)$ for $x<x_0$. 
A similar OPDF has been found for dynamic MTs in 
the presence of MT end-tracking molecular motors \cite{Tischer2010}.

With  increasing length $x$, also the force $f(x)$ increases and, thus, 
$v_{+}[f(x)]$ decreases and $\omega_{c}[f(x)]$ grows exponentially. 
If $x$ becomes sufficiently large that the condition $\lambda^{-1}[f(x)]<0$ 
holds, the distribution $P(x)$ starts to {\em decrease}
exponentially. In this length regime the MT undergoes a catastrophe with high
probability.
Because the distribution always decreases exponentially for sufficiently 
large $x$,  a single MT growing against an elastic
obstacle is always in the regime of bounded growth regardless of how large 
the values of $\omega_{\text{on}}$ and $\omega_{r}$ are chosen. 
This behavior is a result of the linearly increasing force,
which  gives rise to arbitrarily large forces for increasing  $x$
 in contrast to  growth
under constant or zero force, 
where a MT can either be in a phase of bounded or
unbounded growth as mentioned above. 

\begin{figure*}
 \includegraphics[width=\textwidth]{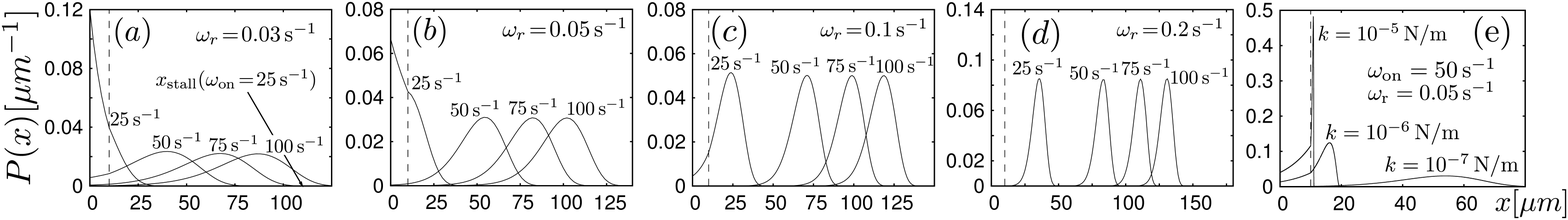}
 \caption{
Stationary MT length distribution 
  $P(x)$ in the steady state  for growth against an elastic obstacle 
  with $\omega_{\text{on}}=25 \,{\rm
     s}^{-1}, 50\,{\rm s}^{-1}, 75\,{\rm s}^{-1}, 100\,{\rm s}^{-1}$ and
   different values of $\omega_{r}$. We set $k=10^{-7}\,{\rm N/m}$ and
   $x_{0}=10^{-5}\, {\rm m}$. (a) $\omega_{r}=0.03\,{\rm s}^{-1}$. 
  (b) $\omega_{r}=0.05
   \,{\rm s}^{-1}$. (c) $\omega_{r}=0.1 \,{\rm s}^{-1}$. (d) $\omega_{r}=0.2
   \,{\rm s}^{-1}$. Dashed line represents $x_{0}$. In picture (a) the stall
   length $x_{s}$ for $\omega_{\text{on}}=25\,{\rm s}^{-1}$, obtained from
   simple polymerization kinetics, is indicated by an
   arrow.
 (e): $P(x)$ for $\omega_{\text{on}}=50 \,{\rm
     s}^{-1}$, $\omega_{r}=0.05 \,{\rm s}^{-1}$ and different values 
  of the spring constant $k$.
  \label{fig:P(x)}}
\end{figure*}

The behavior is also in contrast to length distributions in 
confinement between
fixed rigid walls, where we found a 
 transition between exponentially decreasing and increasing 
length distributions: The elastic obstacle typically 
leads to a non-monotonic length distribution with a {\em maximum} 
in the region $x>x_0$.
(as long as the on-rate  $\omega_{\text{on}}$ and rescue rate 
$\omega_{r}$ are sufficiently large and 
the obstacle stiffness $k$ sufficiently small). 
 While rescue events (and an exponential decrease in the growth 
velocity $v_+[f(x)]$)  cause $P(x)$ to
increase exponentially for small MT length, catastrophes are responsible for
an exponential decrease for large $x$. The interplay between rescues and
catastrophes gives rise to strongly {\em localized} 
probability distributions with a maximum.
Figs.\ \ref{fig:P(x)} (a-d)
show the steady state distribution $P(x)$ obtained from
eq.\ (\ref{eq:Pf}) for different values of $\omega_{\text{on}}$ and
$\omega_{r}$. We chose $k=10^{-7}\,{\rm N/m}$ and $x_{0}=10\,{\rm \mu m}$. 
In the steady state, a
stable length distribution with a well defined average length $\langle x
\rangle=\int _{0}^{\infty} P(x)xdx$ is maintained although the MT is still
subject to dynamic instability. 
The length distributions drop to zero for large 
$x$, where 
$\lambda^{-1}(x)\!\sim\!-\omega_{c}(x)/v_{+}(x)$ and $\omega_{c}(x)/v_{+}(x)$
increases exponentially with increasing force.

The most probable MT length $x_{mp}$ maximizes 
 the stationary length distribution   (\ref{eq:Pf}).
Because $v_- \gg v_+(x)$ and 
using the approximation of an exponentially decreasing  growth velocity,
 $v_{+}[f(x)]\approx
v_{+}(0)e^{-f(x)}$, which is valid for $\omega_{\text{on}}\gg \omega_{\text{off}}$
(see eq.\ (\ref{eq:vel_f})),
we obtain a condition $\lambda^{-1}(x_{mp})  =- \partial_xf(x_{mp}) =- k/F_0$
or 
\begin{equation}
\label{eq:condition_fmp}
 v_{+}(f_{mp})\omega_{r} -v_{-}\omega_{c}(f_{mp}) = -(k/F_0 ) v_- v_+(f_{mp})
\end{equation}
for the corresponding most probable force $f_{mp} = (k/F_0)(x_{mp}-x_0)$.

For  an 
exponentially increasing 
catastrophe rate above the  characteristic force $F_0$,
eq.\ (\ref{eq:wcat_app}), we find 
\begin{equation}
\label{eq:fmp}
   f_{mp} \sim  \frac{1}{2}\ln\left[
\frac{v_+(0) \omega_r}{v_- \omega_c(0)}
  \left( 1+ \frac{k v_-}{F_0\omega_r}\right) \right]
\end{equation}
We can distinguish two limits:
(i) For a {\em soft} obstacle with $kv_-/F_0\omega_r \ll 1$ 
the most probable force $f_{mp}$ is identical to the   critical 
force $f_c$ for MT dynamics under constant force, see (\ref{eq:fc}), 
 because the  right hand side in the condition (\ref{eq:condition_fmp})
for $f_{mp}$ 
can be neglected and we exactly recover 
condition (\ref{eq:condition_fc}) for $f_c$. 
The most probable 
MT length thus ``self-organizes'' into a  ``critical'' state 
with  $f_{mp}\approx f_c$, and 
a MT  pushing against a soft  elastic obstacle  generates
the same force as if growing against  a constant force.
This  force grows logarithmically in the on-rate $\omega_{\text{on}}$
{\em and} the rescue rate $\omega_r$.
(ii) For a {\em stiff}  obstacle with $kv_-/F_0\omega_r \gg 1$, 
on the other hand, 
the most probable force is larger than the critical force, 
$f_{mp} \gg f_c$, and the MT
growing against a stiff obstacle generates a {\em higher} force.
This limit can also be realized for  vanishing rescue rate $\omega_r$,
and for  $kv_-/F_0\omega_r \gg 1$ we indeed 
recover the maximal pushing force in the absence of rescue events, 
i.e.\  $f_{mp} \approx \langle f_{\text{max}} \rangle$ 
from eq.\ (\ref{eq:w_r0_f_max2}) with $v_+(0)\approx
\omega_{\text{on}}d$.  This  force
 grows logarithmically in the on-rate $\omega_{\text{on}}$.
Furthermore, if $f_{mp}$ becomes  negative   for  small on-rates  
and rescue rates  (leading to $\lambda^{-1}(0)<-k/F_0$,
 see eq.\ (\ref{eq:fmp})) 
the stationary length distribution has no maximum,
see  for example Figs.\ \ref{fig:P(x)}(a,b) at the lowest on-rates.

With respect to the MT's ability to generate force 
the two limits can be interpreted also in the following way:  
$F_0$ is the characteristic force above which the catastrophe rate 
increases exponentially. 
For $kv_-/F_0\omega_r \ll  1$, the average length loss
during a period of shrinkage, $v_-/\omega_r$, is much smaller than 
the length $F_0/k$, which is the displacement $x-x_0$
of the elastic obstacle 
under the characteristic force $F_0$. Therefore, the MT tip always remains 
in the region $x>x_0$ under the influence of the force
for a soft obstacle with   $kv_-/F_0\omega_r \ll 1$, 
whereas it typically shrinks back into the force-free region $x<x_0$
before the next  rescue event for a stiff obstacle  $kv_-/F_0\omega_r \gg 1$.
The force generation by the MT can only be enhanced by rescue events
if rescue takes place under force in the regime $x>x_0$. Therefore, 
we find an increased polymerization force 
  $f_{mp}  \approx f_c \gg \langle f_{\text{max}} \rangle$ 
as compared to the force $f_{\text{max}}$ 
 without rescue events discussed in the 
previous section only in the limit 
$kv_-/F_0\omega_r \ll  1$, i.e., for a soft 
obstacle or sufficiently large rescue rate. 
In the limit $kv_-/F_0\omega_r \gg  1$ of a stiff obstacle, the MT 
only generates the same force as in the absence of rescues,
$f_{mp}  \approx  \langle f_{\text{max}} \rangle$.

By comparing the condition (\ref{eq:condition_fc}) or 
$v_+(f_c) = v_-\omega_c(f_c)/\omega_r$ for the 
critical force $f_c$,  the condition   (\ref{eq:condition_fmp})
or $v_+(f_{mp}) = v_-\omega_c(f_{mp})/\omega_r(1+kv_-/F_0)< 
v_-\omega_c(f_{mp})/\omega_r$  for the most probable force $f_{mp}$, and 
the condition $v_+(f_{\text{stall}}) =0$ for the stall force, see 
eq.\ (\ref{eq:fstall}), it 
follows that 
\begin{equation}
   f_{c} \le f_{mp} \ll  f_{\text{stall}}
\label{eq:force_order}
\end{equation}
i.e., force generated against an elastic obstacle 
is between critical and stall force but typically well 
below the stall force, which is  
the maximal polymerization force in the 
absence of catastrophes.
Therefore, the  stall length
$x_{\rm stall}=(F_{0}/k)\ln \left( \omega_{\text{on}}/\omega_{\text{off}} \right)
+x_{0}$  is always much larger than the most probable 
 MT length $x_{mp}$ at the maximum of the  stationary length distribution, 
see Fig.\ \ref{fig:P(x)}(a). 
This shows that the dynamic instability reduces the typical MT  length
significantly compared to simple polymerization kinetics.

In order to quantify the width of the stationary distribution $P(x)$
we expand the exponential in  (\ref{eq:Pf}) 
up to second order about the maximum at  
$x_{mp}$. 
To do so we first expand $\lambda^{-1}(x)$  up to 
first  order:
\begin{equation}\label{eq:approx_lambda}
  \lambda^{-1}(x)\approx -\frac{k}{F_{0}}
  \left[  \frac{v_{+}(x_{mp})\omega_{r}+  
   v_{-}\omega_{c}(x_{mp})}
        {v_{+}(x_{mp})v_{-} } \right] 
       \left( x-x_{mp} \right)
\end{equation}
where we used  $v_{+}[f(x)]\approx
v_{+}(0)e^{-f(x)}$, which is valid  for 
 $\omega_{\text{on}}\gg \omega_{\text{off}}$
(see eq.\ (\ref{eq:vel_f})), and where we 
 approximated the catastrophe rate by an exponential 
$\omega_c[f(x)]\approx \omega_c(0)e^{f(x)}$ 
according to eq.\ (\ref{eq:wcat_app})
resulting in $\omega_{c}'[f(x)]\approx k\omega_{c}[f(x)]/F_0$.
The prime denotes a derivative with respect to
the length $x$.
Using  the expansion (\ref{eq:approx_lambda}) in 
eq.\ (\ref{eq:Pf}), 
 we obtain an  approximately Gaussian length distribution  
\begin{align}
 P(x) \approx &\  N \left( 1+\frac{v_{-}}{v_{+}(x)}\right) 
     e^{ x_{0}/\lambda(0)}\ \times \nonumber  \\
         & \exp \left[ \frac{(x_{mp} - x_{0})^2}{2\sigma^2}\right]
               \exp \left[- \frac{(x-x_{mp})^2}{2\sigma^2}\right]
  \label{eq:PF_approx}
\end{align}
with a width 
\begin{eqnarray}
\sigma^{2}&=&\frac{F_{0}}{k}\left[  \frac{v_{+}(x_{mp})v_{-} }
       {v_{+}(x_{mp})\omega_{r}+
        v_{-}\omega_{c}(x_{mp})} \right]
  \nonumber\\
   &\approx& \left(\frac{F_{0}}{k}\right)^2
      \left(1+ \frac{2F_0\omega_{r}}{kv_-} \right)^{-1}
\label{eq:sigmaP}
\end{eqnarray}
where we used the saddle point condition (\ref{eq:condition_fmp}) in  the last
approximation and  the exponential approximations $v_{+}[f(x)]\approx
v_{+}(0)e^{-f(x)}$ and $\omega_c[f(x)]\approx \omega_c(0)e^{f(x)}$.
Again we have to distinguish the two limits of soft and stiff 
obstacles: (i) For a soft obstacle with  $kv_-/F_0\omega_r \ll 1$
we find $\sigma^2 \approx F_0v_-/2k\omega_r$.
This shows that the width of the length 
distribution decreases 
with increasing  $\omega_{r}$ but is roughly independent 
of the on-rate $\omega_{\text{on}}$, as 
 can also be seen in the series of simulation results shown in 
  Figs.\ \ref{fig:P(x)}.
Closer inspection of the 
simulation results shows that the width of the stationary 
length distribution $P(x)$ is  slightly decreasing 
with the  on-rate $\omega_{\text{on}}$.
(ii)  For a stiff obstacle with  $kv_-/F_0\omega_r \gg 1$, on the other hand, 
we find   $\sigma^2 \approx (F_0/k)^2$, which only depends on obstacle 
stiffness.
All in all, $\sigma^2$ is monotonously decreasing for increasing 
stiffness $k$.

For a soft obstacle $kv_-/F_0\omega_r \ll 1$,  
high rescue rates  thus lead to a sharply 
peaked length distribution $P(x)$ and 
 suppress fluctuations of the MT 
length around $x=x_{mp}$ and we expect 
 $\langle x \rangle\approx x_{mp}$ to a very good approximation.
This property of a sharp maximum in $P(x)$ will make the  
 mean field approximation that is discussed in the 
next section very accurate.

If the obstacle stiffness  $k$ is increased the 
 most probable MT length $x_{mp}= x_0+ f_{mp}F_0/k$
approaches $x_0$,
 and a considerable 
probability weight   is shifted to MT lengths $x$ below $x_{0}$ (see Fig.\
\ref{fig:P(x)} (e)). The average length approaches and finally drops below
$x_{0}$. This signals that  the force generated by the MT
is no longer sufficient to push the
obstacle out of its equilibrium position $x_{0}$.
The obstacle now serves as a fixed rigid boundary and $P(x)$ approaches 
the results eq.\
(\ref{eq:conf_q}) and (\ref{eq:conf_norm}). The dynamics of a single MT within
confinement can therefore be seen as a special case of the dynamics in the
presence of an elastic  obstacle, i.e., for small
$\omega_{\text{on}}$ and $\omega_{r}$ or for large spring constants $k$.

\begin{figure}
 \includegraphics[width=\columnwidth]{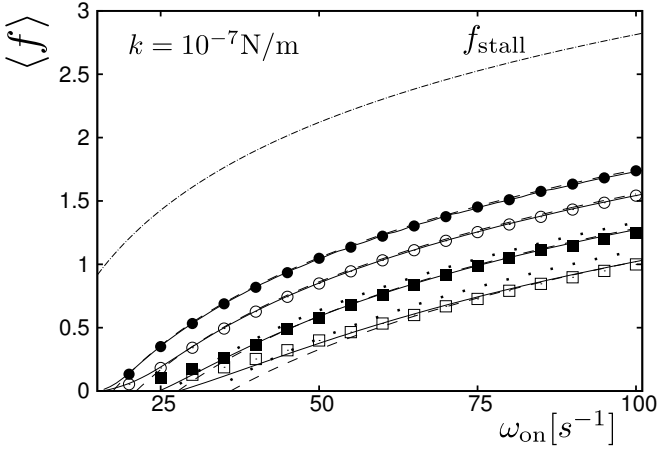}
 \caption{
 Average steady state force $\langle f \rangle$ as a function of
   $\omega_{\text{on}}$  for growth against an elastic obstacle 
  with $\omega_{r}=0.03\,{\rm s}^{-1}(\boxdot),0.05\,{\rm
     s}^{-1}(\blacksquare),0.1\,{\rm s}^{-1}(\odot),0.2\,{\rm
     s}^{-1}(\bullet)$ and $k=10^{-7}\rm{N/m}$. 
  Solid lines: $\langle f \rangle =\int ^{\infty} _{0} f(x)P(x)dx$ with $P(x)$
  given by eqs.\ (\ref{eq:Pf}) and  (\ref{eq:norm_f}).
 Dashed
   lines: $\langle f \rangle$ calculated from mean field 
 equation (\ref{eq:mf_fs}).
   Dotted lines: most probable force $f_{mp}$, measured in simulations, for 
  $\omega_{r}=0.03\,{\rm s}^{-1}$ and $\omega_{r}=0.05\,{\rm s}^{-1}$.
   Also shown is the dimensionless stall force
   $f_{\text{stall}}$ obtained from simple polymerization
   kinetics (\ref{eq:fstall}).
 \label{fig:avg_fs}}
\end{figure}

So far we have quantified the generated force by the most probable force 
$f_{mp}$. The generated force
can also be quantified by  the average steady-state force
$\langle f \rangle =\int ^{\infty} _{0} f(x)P(x)dx$.
Using the stationary distribution (\ref{eq:Pf}) with normalization 
(\ref{eq:norm_f}) we can calculate $\langle f \rangle$;
results are shown  in Fig.\
\ref{fig:avg_fs} in comparison with the most probable force $f_{mp}$,
which is determined numerically from the maximum of $P(x)$,
and the  stall force $f_{\text{stall}}$ 
in the absence of dynamic instability from eq.\ (\ref{eq:fstall}). 
For $\langle f \rangle$, there is  excellent agreement with  stochastic
simulations  over the complete range of parameter
values. The results clearly show that the  dynamic instability reduces
the ability to generate polymerization forces  since, even for large values of
$\omega_{\text{on}}$ and $\omega_{r}$, the average force 
$\langle f \rangle $ is always smaller than the
stall force. Nevertheless forces
up to $F\sim 1.5\,F_{0}$ can be obtained in the steady state for realistic
parameter values. Comparing $\langle f \rangle$ and  $f_{mp}$ we find 
$\langle f \rangle \le f_{mp}$, and both forces become identical, 
$\langle f \rangle \approx  f_{mp}$, in the limit of large rescue rates
or a soft obstacle $kv_-/F_0\omega_r \ll 1$, where also the length
distributions $P(x)$  become sharply peaked, see Fig.\  \ref{fig:P(x)}. 
Comparing different combinations of $\omega_{\text{on}}$ and
$\omega_{r}$ and the corresponding forces, one finds that the influence
of the on-rate 
$\omega_{\text{on}}$ on force generation is more significant than
the influence of the rescue rate $\omega_{r}$. 
For $\omega_{\text{on}}=100\,{\rm s}^{-1}$, a four fold
increase of the rescue rate $\omega_{r}$ gives rise to  an increase of 
$\langle f \rangle $ by a factor of
$\sim 1.5$, while for $\omega_{r}=0.1\,{\rm s}^{-1}$, a four fold
increase of the on-rate  $\omega_{\text{on}}$ results in an amplification of 
the force $\langle f \rangle $ by a factor of $\sim 9$.
These results can be explained within a mean field theory 
presented in the next section. 

\begin{figure}
 \includegraphics[width=\columnwidth]{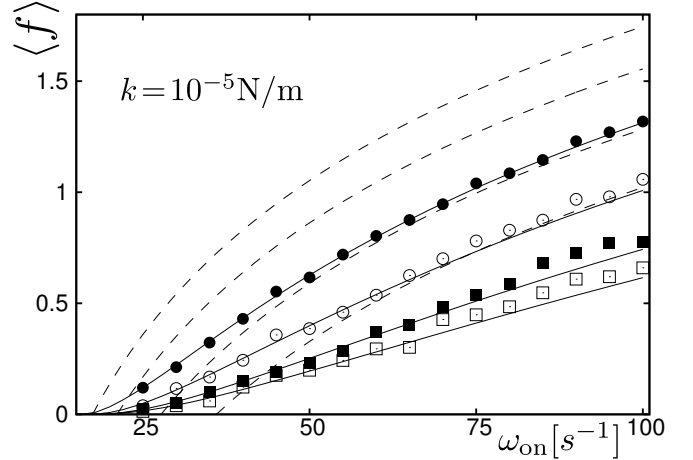}
 \caption{
 Average steady state force $\langle f \rangle$ as a function of
   $\omega_{\text{on}}$  for growth against an elastic obstacle 
  with $\omega_{r}=0.03\,{\rm s}^{-1}(\boxdot),0.05\,{\rm
     s}^{-1}(\blacksquare),0.1\,{\rm s}^{-1}(\odot),0.2\,{\rm
     s}^{-1}(\bullet)$ and $k=10^{-5}\rm{N/m}$. 
 Solid lines: $\langle f \rangle =\int ^{\infty} _{0} f(x)P(x)dx$ with $P(x)$
 given by eqs.\ (\ref{eq:Pf}) and  (\ref{eq:norm_f}).
Dashed lines from bottom to top: 
$\langle f \rangle$ calculated from mean field equation (\ref{eq:mf_fs}) 
for $\omega_{r}=0.03\,{\rm s}^{-1},0.05\,{\rm s}^{-1},0.1\,{\rm s}^{-1}$ 
and $0.2\,{\rm s}^{-1}$.
\label{fig:avg_fs_k1e05}}
\end{figure}

\subsection{Mean field approach (non-zero rescue rate)}

In the following, we show that we can reproduce many of the results for the
average polymerization force $\langle f \rangle$ for non-zero rescue rate
using a simplified mean field approach. Using the mean field approach, we can
also address the time evolution of the average force $\langle f \rangle$, for
example, in dilution experiments.  Since the switching between the two states
of growth is a stochastic process, the length $x$ and the force $f(x)$ are
stochastic variables. Therefore, the velocity of growth $v_{+}[f(x)]$ and
the catastrophe rate $\omega_{+}[f(x)]$ also become stochastic variables which,
in the steady state, fluctuate around their average values.  Within the mean
field approach we neglect these fluctuations and use $\langle v_{+}[f(x)]
\rangle = v_+(\langle f \rangle)$ and $\langle \omega_{+}[f(x)]\rangle =
\omega_+(\langle f \rangle)$.  In the mean field approximation, the average time
in the growing state is given by $1/\omega_{c}(\langle f \rangle)$ and the
average growth velocity is $v_+(\langle f \rangle)$. The average time in a
shrinking state is $1/\omega_r$. Therefore, the mean field probabilities to
find the MT growing or shrinking are $p_+ = \omega_{r}/[\omega_{r} +
\omega_{c}(\langle f \rangle)]$ and $p_- = \omega_{c}(\langle f
\rangle)/[\omega_{r} + \omega_{c}(\langle f \rangle)]$, respectively.  This
results in the following mean field average velocity $v$ of a single MT under
force:
\begin{equation}
\label{eq:mf_vel}
  v(\langle f \rangle )= \frac{v_{+}(\langle f \rangle)\omega_{r} 
      -v_{-}\omega_{c}(\langle f \rangle) }
        {\omega_{r} + \omega_{c}(\langle f \rangle) }.
\end{equation}
In the steady state the barrier is pushed so far that 
$\langle f \rangle$
stalls the MT. We require $v(\langle f \rangle)=0$ and obtain the 
condition 
\begin{equation}\label{eq:mf_fs}
 v_{+}(\langle f \rangle)\omega_{r} =v_{-}\omega_{c}(\langle f \rangle)
\end{equation}
for the stationary state. This condition 
corresponds to a force, where the  average length 
gain during growth, $v_{+}(\langle f \rangle)/ \omega_{c}(\langle f \rangle)$, 
equals the average length loss during shrinking, $v_-/\omega_r$.
From the mean field equation 
 (\ref{eq:mf_fs}), the average steady state force, $\langle f
\rangle$ can be calculated as a function of $\omega_{r}$ and
$\omega_{\text{on}}$. The average length 
$\langle x \rangle$ can be  obtained from the relation
$\langle f \rangle = (k/F_{0})(\langle x \rangle -x_{0})$.
Results obtained from the mean field equation
(\ref{eq:mf_fs}) match numerical results from
stochastic simulations very 
well as shown in Fig.\ \ref{fig:avg_fs}.

The mean field condition (\ref{eq:mf_fs})
 is identical to  the condition (\ref{eq:condition_fc}) for the 
critical force  $f_c$ for MT dynamics under constant force
such that  
\begin{equation}
  \langle f \rangle = f_c,
\label{eq:fav=fc}
\end{equation}
which 
 can be interpreted as ``self-organization'' of  
the average MT length or the average force to the  ``critical'' state.
Therefore, the curves presented in Fig.\ \ref{fig:avg_fs} for 
$\langle f \rangle$ are 
identical to the curves shown in Fig.\ \ref{fig:Fconst} (b) for 
$f_c$. 

This  also allows us to take over the results 
we derived for the  critical constant force $f_c$.
Using the approximation of an exponentially decreasing  growth velocity,
 $v_{+}[f(x)]\approx
v_{+}(0)e^{-f(x)}$, which is valid for $\omega_{\text{on}}\gg \omega_{\text{off}}$
(see eq.\ (\ref{eq:vel_f})), and 
an  exponentially increasing 
catastrophe rate above the  characteristic force $F_0$,
eq.\ (\ref{eq:wcat_app}), we find 
\begin{equation}
\label{eq:fav}
   \langle f \rangle  \sim  \frac{1}{2}\ln\left(
\frac{v_+(0) \omega_r}{v_- \omega_c(0)}\right).
\end{equation}
which is identical to the result (\ref{eq:fc}) for $f_c$. 

Comparing with the stall force and the most probable force, 
we use relation (\ref{eq:force_order}) and find 
\begin{equation}
   \langle f \rangle = f_{c} \le f_{mp} \ll  f_{\text{stall}}.
\label{eq:force_order2}
\end{equation}
In the limit of a soft obstacle, $kv_-/F_0\omega_r \ll 1$, 
the average force $\langle f \rangle$ approaches the 
most probable force $\langle f \rangle \approx f_{mp}$, whereas
the mean field  average force
 $\langle f \rangle$
 is always smaller than the stall force $f_{\text{stall}}$
in the absence of dynamic instability 
from eq.\ (\ref{eq:fstall}).

Finally, we discuss the limits of validity of the mean field 
approximation.
The mean field approximation is based on the existence of a 
pronounced maximum in the stationary MT length distribution $P(x)$,
which contains most of the weight of the probability density $P(x)$. 
It breaks down if this  maximum
broadens or vanishes, such that a  considerable
amount of probability density is shifted below $x_{0}$ into the regime 
of force-free growth. 
Then the MT typically 
shrinks into the force-free region $x<x_0$ during 
phases of shrinkage such that  the growing phase explores the whole 
range of forces starting from $f=0$ up to $f> \langle f \rangle$,
and the approximation of a constant  
average force $f \approx \langle f \rangle$  during growth is no longer 
fulfilled. 
For small spring constants $k$ or large values of
 $\omega_{r}$, the length distribution 
$P(x)$ assumes a Gaussian shape 
with width $\sigma$, see eqs.\ (\ref{eq:PF_approx}) and (\ref{eq:sigmaP}). 
When $k$ is increased for a
fixed combination of $\omega_{\text{on}}$ and $\omega_{r}$, the average length
$\langle x \rangle$ approaches $x_{0}$ as $\langle x \rangle-x_0 \propto 1/k$,
whereas the  width $\sigma$ of the length distribution 
only decreases as $\sigma \propto 1/\sqrt{k}$ in the regime of a 
soft obstacle $kv_-/F_0\omega_r \ll 1$, as can be seen from eq.\
(\ref{eq:sigmaP}). Therefore, an increasing
amount of probability density is shifted below $x_{0}$, where no force is
acting on the MT ensemble (see Figs.\ \ref{fig:P(x)})(a) and (e)). 
The mean
field approximation is only valid for spring constants $k$ which fulfill
$\langle x \rangle -x_{0}\gg \sigma/2$ for given parameters
$\omega_{\text{on}}$ and $\omega_{r}$.
With  $\langle f \rangle =(k/F_{0})(\langle x \rangle -x_{0})$ this is
equivalent to a condition 
\begin{equation}
 \langle f \rangle \gg \frac{k\sigma}{2F_0} \approx  
   \frac{1}{2}
      \left(1+ \frac{2F_0\omega_{r}}{kv_-} \right)^{-1/2}
\end{equation}
according to eq.\ (\ref{eq:sigmaP}). 
This condition can only be fulfilled in the limit of 
 a {\em soft} obstacle with $kv_-/F_0\omega_r \ll 1$.
For the validity of the mean field approximation 
we therefore  recover the condition  that the average length loss
during a period of shrinkage, $v_-/\omega_r$, is much smaller than 
the typical displacement  $F_0/k$ of the elastic obstacle 
under the characteristic force $F_0$. Then the MT tip always remains 
in the region $x>x_0$ under the influence of the force.

\subsection{Dynamics and dilution experiments}

\begin{figure}
 \includegraphics[width=\columnwidth]{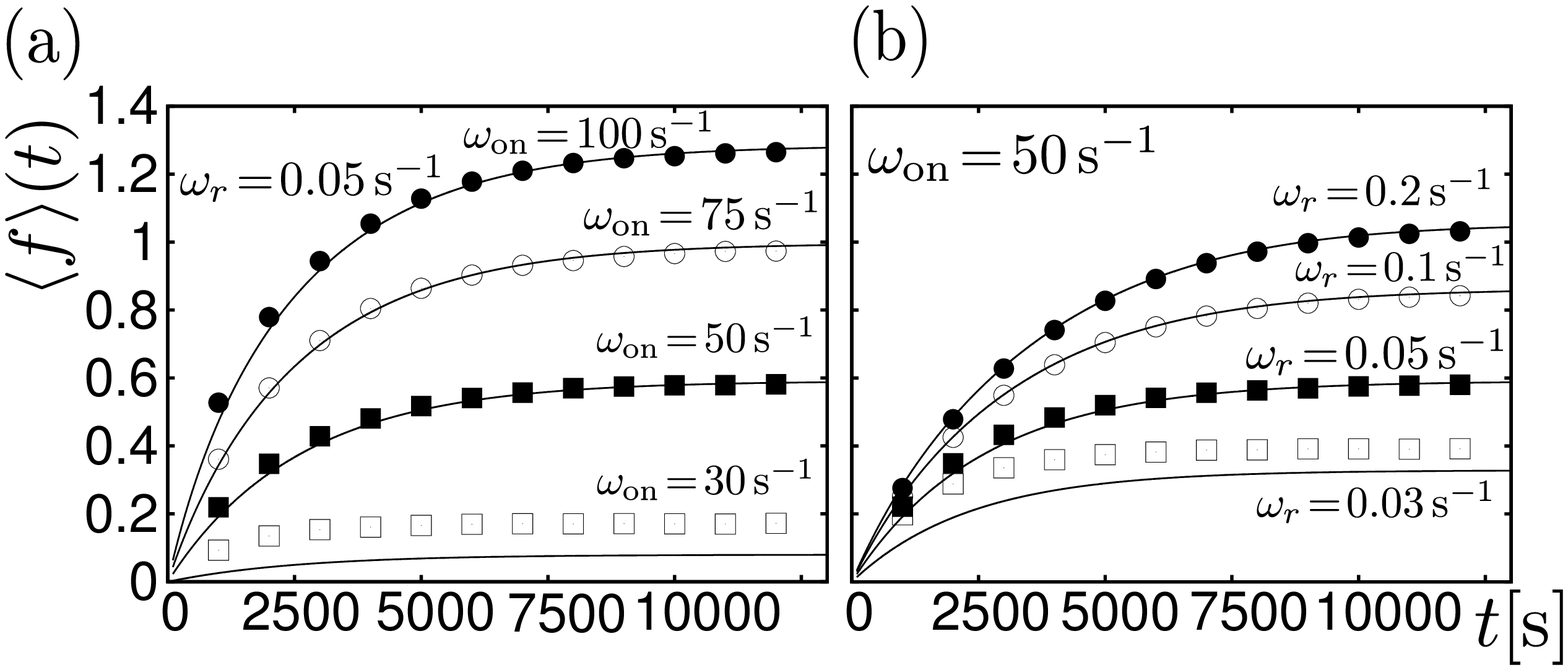}
\caption{ (a): Average force $\langle f \rangle(t)$ as a function of time for
  $k=10^{-7}\,{\rm N/m}$, $\omega_{r}=0.05\,{\rm s}^{-1}$ and different values
  of $\omega_{\text{on}}$.  Symbols: time dependent average force $\langle f
  \rangle(t)$ measured in simulations. Solid lines: time dependent average
  force trajectory calculated from eq.\ (\ref{eq:time_evo}).  
(b): Average force $\langle f
  \rangle(t)$ as a function of time for $k=10^{-7}\,{\rm N/m}$,
  $\omega_{\text{on}}=50\,{\rm s}^{-1}$ and different values of $\omega_{r}$.
  Symbols: time dependent average force $\langle f \rangle(t)$ measured in
  simulations. Solid lines: time dependent average force trajectory calculated
  from eq.\ (\ref{eq:time_evo}).\label{fig:mf_time_evo}}  
\end{figure}

Within the mean field approach we can also derive 
 an analytical time evolution of the
average time-dependent force $\langle f \rangle(t)$. 
The time evolution is based on  eq.\
(\ref{eq:mf_vel}), which gives 
 a mean field approximation for the average MT velocity 
$v(\langle f \rangle )$  as a  function of the average force. 
On the other hand, the average MT growth velocity is related to the 
time derivative of the average  force by 
\begin{equation}
\frac{d}{dt} \langle f \rangle = \frac{k}{F_0}\frac{d}{dt} \langle x \rangle
=  \frac{k}{F_0}  v(\langle f \rangle)
\label{eq:time_evo}
\end{equation}
Using eq.\ (\ref{eq:mf_vel}) for $v(\langle f \rangle )$, 
this gives a mean field  equation of motion for $\langle f \rangle$
similar to  eq.\ (\ref{eq:time_evo0}) in the absence of rescue 
events. 
Integrating this equation numerically we obtain mean field 
trajectories  for the average  force $\langle f \rangle(t)$ as a
function of time $t$.  
Figs.\ \ref{fig:mf_time_evo} shows such 
trajectories for  $k=10^{-7}{\rm N/m}$ and a initial condition 
$\langle f \rangle(0)=0$ at $t=0$.   
Also shown in Figs.\ \ref{fig:mf_time_evo} are results from 
 stochastic simulations, which show excellent agreement with the 
mean field trajectories.

\begin{figure}
 \includegraphics[width=\columnwidth]{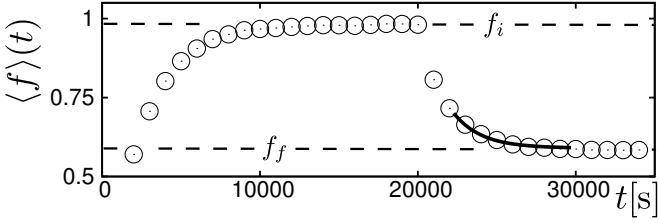}
 \caption{Average force $\langle f \rangle(t)$ as a function of time
   $t$. Symbols are results obtained from simulations. We set
   $k=10^{-7}\,\rm{N/m}$, $\omega_{r}=0.05\, \rm{s}^{-1}$ and
   $\omega_{\text{on}}=75\,\rm{s}^{-1}$. At $t=20000\,\rm{s}$,
   $\omega_{\text{on}}$ is diluted down to $\omega_{\text{on}}=50\,\rm{s}^{-1}$.
Solid line represents a fit with an exponential decay 
   (\ref{eq:time_evo_exp}) to the simulated data with 
   fit parameter $\tau_d = 1762\, \rm{s}$.
Dashed lines indicate the average force in the initial 
state $f_{i}$ before dilution and in the new final 
state $f_{f}$ after dilution.
 \label{fig:mf_dilution}}
\end{figure}

We now  address the question of how fast a single MT responds to
external changes of one of its growth parameters. Here we focus on fast
dilution of the tubulin concentration, which is directly related to the
tubulin on-rate $\omega_{\text{on}}$. {\it In vivo} tubulin concentration can
be changed by tubulin binding proteins like stathmin \cite{Curmi1997}, 
while in {\it in
  vitro} experiments, the tubulin concentration can be diluted within seconds
\cite{Walker1991}.
In the following we give a mean field estimate of the typical time scale,
which governs the return dynamics of the MT back to a new steady state after
the tubulin on-rate is suddenly decreased. 
In the initial steady state the average
velocity $v(\langle f \rangle_i)$ vanishes and 
the average polymerization force
$\langle f \rangle_i$ (and, thus, the average length $\langle x \rangle_i$) can
be calculated from the condition 
$v_{+}(\langle f\rangle_i)\omega_{r}=v_{-}\omega_{c}(\langle f \rangle_i)$, 
cf.\ eq.\ (\ref{eq:mf_fs}),
for a given combination of $\omega_{\text{on}}$ and $\omega_{r}$. If
$\omega_{\text{on}}$ is suddenly decreased
this leads to a sudden decrease in the
 growth velocity to $\tilde{v}_{+}(f)< v_{+}(f)$ and an increase of the 
catastrophe rate to $\tilde{\omega}_{c}(f)>\omega_{c}(f)$, resulting in  a
negative average velocity $v(\langle f \rangle)=[\tilde{v}_{+}(\langle f
\rangle)\omega_{r}-v_{-}\tilde{\omega}_{c}(\langle f \rangle)]/[\omega_{r}
+\tilde{\omega}_{c}(\langle f \rangle)]<0$ according to eq.\ (\ref{eq:mf_vel}).
Consequently,  the MT starts to shrink with an 
average  velocity $v(\langle f \rangle)<0$.
This relaxes the force from  the elastic obstacle, i.e., 
 $\langle f \rangle(t)$ starts to decrease from the initial 
value $f_{i} \equiv \langle f \rangle_i$. With decreasing average
force $\langle f \rangle(t)$, the average growth velocity 
$v(\langle f \rangle(t))$ increases  again (because $\tilde{v}_+$ increases and
$\tilde{\omega}_c$ decreases)  until the steady state condition
$\tilde{v}_{+}(\langle f \rangle_f)\omega_{r}=
    v_{-}\tilde{\omega}_{c}(\langle f \rangle_f)$ holds
again and a new steady state force $\langle f \rangle_f<\langle f \rangle_i$ 
is reached (s. Fig.\ \ref{fig:mf_dilution}).

The relaxation dynamics to the new steady state after tubulin dilution
 is therefore governed by the average velocity 
$v(\langle f \rangle)$ given by eq.\ (\ref{eq:mf_vel}).
To extract  a characteristic relaxation time scale, 
we expand the average velocity $v(\langle f \rangle)$ 
to first  order around the final 
steady-state  polymerization force $f_f\equiv \langle f \rangle_f$, 
which is the
solution of eq.\ (\ref{eq:mf_fs}) with $\omega_{r}$ and the decreased tubulin
on-rate $\omega_{\text{on}}$, which takes its dilution value. 
Using $v(f_f)=0$ one finds in first order
\begin{eqnarray}
  v(\langle f \rangle) 
    &\approx& -\left[\frac{v_{+}(f_f)\omega_{r}+v_{-}\omega_{c}'(f_f)}
    {\omega_{r}+\omega_{c}(f_f)} \right]\left( \langle f \rangle -f_f \right)
\label{eq:mf_v_approx}
\end{eqnarray}
where the prime denotes the derivative with respect to the
 force.
In the last approximation we used  the 
mean field condition eq.\ (\ref{eq:mf_fs}) and 
 $v_{+}[f(x)]\approx
v_{+}(0)e^{-f(x)}$, which is valid  for 
 $\omega_{\text{on}}\gg \omega_{\text{off}}$
(see eq.\ (\ref{eq:vel_f})).
This expansion is only valid for average forces close to the new average
polymerization force $f_f$.
Using this expansion, the time evolution  (\ref{eq:time_evo}) of the 
average force after dilution exhibits
an exponential decay
\begin{equation}
     \label{eq:time_evo_exp}
    \langle f \rangle(t) = f_f + (f_i-f_f) e^{-t/\tau_d}
\end{equation}
with a characteristic dilution  time scale
\begin{equation}\label{eq:time_evo_tau}
\tau_d= 
  \frac{F_{0}}{k} \frac{\omega_{r}+\omega_{c}(f_f)} 
    {v_{+}(f_f)\omega_{r}+v_{-}\omega_{c}'(f_f)}
\approx \frac{F_{0}}{k} \frac{\omega_{r}+\omega_{c}(f_f)} 
    {2v_{-}\omega_{c}(f_f)}
\end{equation}
where  we  approximated the catastrophe rate by an exponential 
$\omega_c[f(x)]\approx \omega_c(0)e^{f(x)}$ 
according to eq.\ (\ref{eq:wcat_app}),
and we used the mean field condition eq.\ (\ref{eq:mf_fs}).
In the limit $\omega_c(f_f) \gg \omega_r$, i.e., 
  at forces $f_f \gg 1$, we obtain 
the simple result $\tau_d \approx F_0/2v_-k$. 
In general, the relaxation time $\tau_d$ is proportional to the square
$\sigma^2$ of the width 
of the stationary distribution, cf.\ eq.\ (\ref{eq:sigmaP}): A narrow 
length distribution gives rise to fast relaxation to the new 
average force.

\section{Experimental catastrophe model}\label{sec:exp_wcat}

So far we have employed 
the catastrophe rate derived by Flyvbjerg {\it et al.}, to 
which we will refer as $\omega_{c,\text{Flyv}}$ in the following. 
This expression for the catastrophe rate 
 was based on theoretical calculations of the
inverse passage time to a state with a vanishing GTP-cap, see eq.\
(\ref{eq:wcat}). 
In order to investigate the robustness of our results with respect to 
changes of the catastrophe model, we now investigate an alternative 
expression for the  catastrophe rate that has been obtained 
from experimental results. Throughout this section,  we
focus on the third confinement scenario of an  elastic obstacle,
 and we compare results from the two different  catastrophe models for 
zero rescue rate 
$\omega_{r}=0$ and non-zero rescue rate
$\omega_{r}> 0$. In addition, we restrict the comparison
to mean field results, since numerical and stochastic calculations match
mean field results well over the complete range of parameters
(see Sec. \ref{sec:elastic}).

\begin{figure}
\includegraphics[width=\columnwidth]{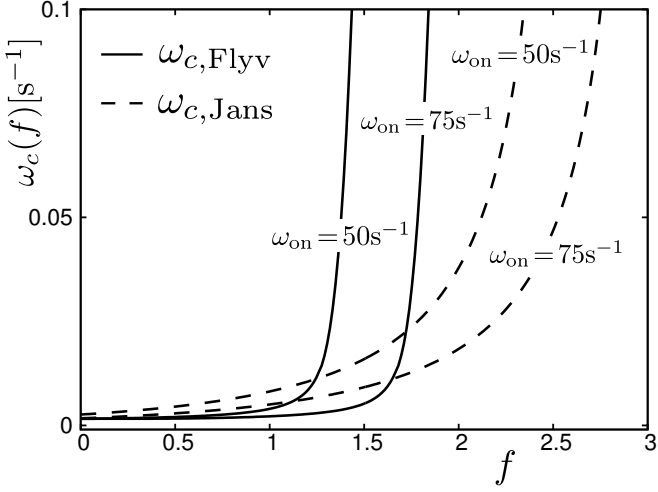}
\caption{ 
 Catastrophe rate $\omega_{c}(f)$ as a function of force $f$ for
  $\omega_{\text{on}}= 50\;\rm{s}^{-1}$ and $\omega_{\text{on}}=
  75\;\rm{s}^{-1}$. Solid lines: $\omega_{c,\text{Flyv}}$ from the catastrophe 
  model by Flyvbjerg {\it et al.}
   Dashed lines: $\omega_{c,\text{Jans}}$ from the experimental catastrophe 
  model by Janson {\it et al}.}
\label{fig:wcat_compare}
\end{figure}

Experimentally, it has been found 
 that the average time $\langle \tau_{+}\rangle$
spent in a growing state is a linear function of the growth velocity $v_{+}$
\cite{Janson2003}. 
The force-dependent catastrophe rate is then given by
\begin{equation}
\label{eq:exp_wcat}
\omega_{c,\text{Jans}}(f)=\frac{1}{av_{+}(f) +b}
\end{equation}
with constant coefficients 
$a=1.38\cdot 10^{10}\, {\rm s}^{2}{\rm m}^{-1}$ and $b=20\; {\rm s}$.
At $v_{+}(f)=0$, $\omega_{c,\text{Jans}}(f)=0.05\,{\rm s}^{-1}$ 
and for $v_{+}(f)=-b/a$, the
catastrophe rate $\omega_{c,\text{Jans}}(f)$ diverges. This is in  contrast to
the theoretical model, where $\omega_{c,\text{Flyv}}(f)$ is finite for all
$v_{+}(f)$. Also $\omega_{c,\text{Jans}}(f)$ increases exponentially for forces
$F>F_{0}$ or $f>1$. This common feature is essential and 
lead to similar results for both catastrophe models. 
In Fig.\ \ref{fig:wcat_compare}, both catastrophe rates are shown
as a function of the dimensionless  force $f$.
The catastrophe model (\ref{eq:exp_wcat}) is based on experimental 
data and, thus, is phenomenological. It assumes neither 
 a purely chemical model, as in the model  by Flyvbjerg {\it et al.},
nor a chemo-mechanical model in the 
sense of ``structural plasticity'' \cite{Kueh2009}.

\subsection{Vanishing rescue rate}

\begin{figure}
\includegraphics[width=\columnwidth]{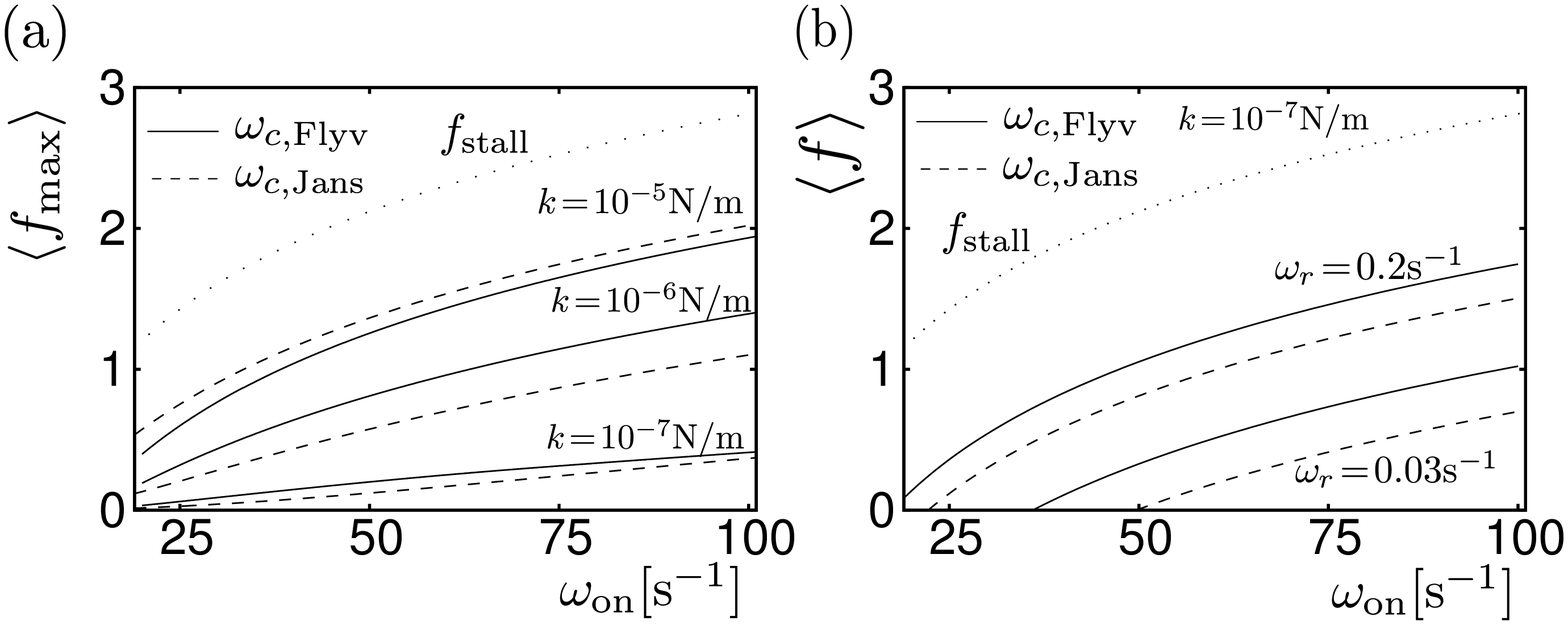}
\caption{ (a): Average maximal polymerization force $\langle f_{\text{max}} \rangle$
  as a function of $\omega_{\text{on}}$ and $\omega_{r}=0$ for
  $k=10^{-7}\,{\rm N/m}$, $10^{-6}\,{\rm N/m}$ and $10^{-7}\,{\rm N/m}$ (top
  to bottom) . Dotted line: dimensionless stall force
  $f_{\text{stall}}$. Solid lines: $\langle f_{\text{max}} \rangle$ obtained
  from $\omega_{c,\text{Flyv}}$ (eq.\ (\ref{eq:w_r0_f_max})). Dashed lines:
  $\langle f_{\text{max}} \rangle$ obtained from $\omega_{c,\text{Jans}}$ (eq.\
  (\ref{eq:exp_wcat_wr0_fmax})).  (b): Average steady state force $\langle
  f\rangle$ as a function of $\omega_{\text{on}}$. $k=10^{-7}\,{\rm N/m}$,
  $\omega_{r}=0.2\,{\rm s}^{-1}$ (top) and $\omega_{r}=0.03\,{\rm
    s}^{-1}$(bottom). Solid lines: $\langle f\rangle$ obtained from
  $\omega_{c,\text{Flyv}}$ (eq.\ (\ref{eq:mf_fs})). Dashed lines: $\langle
  f\rangle$ obtained from $\omega_{c,\text{Jans}}$ (eq.\
  (\ref{eq:exp_wcat_wr0_fs})). Dotted line: dimensionless stall force
  $f_{\text{stall}}$.
\label{fig:exp_wcat_f}}
\end{figure}

We start with the case $\omega_r=0$ without rescue events, 
 and we calculate the average maximal polymerization
force within the experimental catastrophe model  using the self-consistent 
mean field  eq.\  (\ref{eq:w_r0_f_max}), which holds independently 
of the choice of catastrophe model (see Sec.\ \ref{sec:elast_wres0}).  
As for the catastrophe by  Flyvbjerg {\it et al.}, we have
$\omega_{c,\text{Jans}}\tau  \gg 1$ for realistic parameter values and 
$v_{+}(\langle f \rangle)<-b/a$, 
and eq.\ (\ref{eq:w_r0_f_max}) can be solved
explicitly for $\langle f_{\text{max}}\rangle$ in this limit. 
We find an  average maximal
polymerization force 
\begin{equation}\label{eq:exp_wcat_wr0_fmax}
\langle f_{\text{max}}\rangle\approx \ln \left( \left[ 
  \left(A^2+B\right)^{1/2} -A \right] \right)
\end{equation}
with
\begin{align}
A&\equiv \frac{ (\omega_{\text{on}}/\omega_{\text{off}}-1)ad\omega_{\text{off}} 
-  (\omega_{\text{on}}/\omega_{\text{off}}-1)b -\tau}{2\tau}   \nonumber \\ 
B&\equiv 
   \frac{(\omega_{\text{on}}/\omega_{\text{off}}-1)ad\omega_{\text{on}}}{\tau}    
\nonumber.
\end{align}
Since $\omega_{\text{on}}/\omega_{\text{off}} \gg 1$, 
eq.\ (\ref{eq:exp_wcat_wr0_fmax}) can be approximated by
\begin{equation}
\label{eq:exp_wcat_wr0_fmax_approx}
 \langle f_{\text{max}}\rangle\approx 
          \ln \left( \omega_{\text{on}}/\omega_{\text{max}} \right)
\end{equation}
with 
\begin{equation}
  \omega_{\text{max}}\equiv   \frac{2\tau \omega_{\text{off}}}
  { \left[ \left(ad\omega_{\text{off}}-b\right)^{2} 
      + 4ad\omega_{\text{off}}\tau \right]^{1/2} 
    - \left[ad\omega_{\text{off}}-b\right] }
\end{equation}
For realistic parameter values, we have 
$\tau \gg ad\omega_{\text{off}} \ge b$, and recover 
the expression  (\ref{eq:w_r0_f_max2})   derived using 
the Flyvbjerg catastrophe model:
\begin{equation}\label{eq:exp_wcat_wr0_fmax_approx2}
 \langle f_{\text{max}}\rangle\approx 
         \frac{1}{2} 
          \ln \left( \frac{\omega_{\text{on}}^2ad}{\omega_{\text{off}}\tau} \right)
    \approx \frac{1}{2} 
      \ln \left( \frac{\omega_{\text{on}}dk}{F_0\omega_{c,\text{Jans}}(0)} \right).
\end{equation}
In Fig.\ \ref{fig:exp_wcat_f} (a), 
 $\langle f_{\text{max}}\rangle$ as  obtained from 
eq.\ (\ref{eq:w_r0_f_max}) with the Flyvbjerg catastrophe model  
and eq.\ (\ref{eq:exp_wcat_wr0_fmax}) with the experimental 
 catastrophe model  are  shown 
as a function of $\omega_{\text{on}}$. Results match qualitatively and
quantitatively well, although they are obtained from two different catastrophe
models.  The maximal polymerization force 
$\langle f_{\text{max}}\rangle$ always remains smaller than the stall 
 force $f_{\text{stall}}$.

\subsection{Non-zero rescue rate}

Now we  compare both catastrophe models for a non-zero rescue rate, and we
calculate the average steady state force. For the experimental catastrophe 
rate  (\ref{eq:exp_wcat}), 
the mean field equation
(\ref{eq:mf_fs}) can be solved explicitly, 
and the average steady-state force
$\langle f\rangle$ is given by
\begin{equation}\label{eq:exp_wcat_wr0_fs}
 \langle f\rangle=\ln \left( \omega_{\text{on}}/\omega_{\text{av}} \right),
\end{equation}
with 
\begin{equation}
 \omega_{\text{av}}\equiv 
  \left( \left(\frac{b}{2ad}\right)^{2} + 
       \frac{v_{-}}{\omega_{r}ad^2}\right)^{1/2}-\frac{b}{2ad} +\omega_{\text{off}}
\end{equation}
Again $\langle f\rangle < f_{\text{stall}}$ since
$\omega_{\text{av}}>\omega_{\text{off}}$. Fig.\ \ref{fig:exp_wcat_f} (b)
show $\langle f\rangle$ as a function of $\omega_{\text{on}}$.
For  realistic parameter values, we have 
$v_-/\omega_r \gg b^2/a$ and $(v_-/\omega_rad^2)^{1/2}\gg
\omega_{\text{off}}$, 
and recover the expression  (\ref{eq:fav})   derived using 
the Flyvbjerg catastrophe model:
\begin{equation}
 \langle f\rangle\approx 
   \frac{1}{2}\ln \left( \frac{\omega_{\text{on}}^2\omega_rad^2}{v_-}\right)
   \approx 
 \frac{1}{2}\ln \left( \frac{v_+(0)\omega_r}{v_-\omega_{c,\text{Jans}}(0)}\right).
\end{equation}
In Fig.\ \ref{fig:exp_wcat_f} (b), 
results for $\langle f\rangle$ 
from both catastrophe models are shown as a function of on-rate 
$\omega_{\text{on}}$. The average steady state force obtained from
$\omega_{c,\text{Flyv}}$ is always slightly 
larger than $\langle f\rangle$ obtained
from $\omega_{c,\text{Jans}}$, since
$\omega_{c,\text{Jans}}(f)>\omega_{c,\text{Flyv}}(f)$ for forces smaller than or
comparable to $F_{0}$. Otherwise, both results agree qualitatively and 
quantitatively well.

\section{Force-velocity relation}\label{sec:force-velocity-theta}

Finally, we  discuss the influence of the force-velocity relation on the MT
dynamics.
We restrict our analysis to mean field results obtained 
for the third scenario, i.e., the elastic obstacle.
A change in the force-velocity relation directly modifies the
velocity of growth $v_{+}(f)$, but it also affects the catastrophe rate
$\omega_{c}(v_{+}(f))$, which are both crucial parts of the MT dynamics. In
the following, we employ  a more general form of the 
force-velocity relation, which is consistent with thermodynamic constraints, 
 and we show
that our results are robust with respect to this generalization.

In their investigation of experimental data Kolomeisky {\it et al.} 
used  a generalized  growth velocity
\begin{equation}
 v_{+}(f,\theta)=d\{\omega_{\text{on}}\exp(-\theta f) 
    - \omega_{\text{off}}\exp[(1-\theta)f]\},
\label{eq:vel_theta}
\end{equation}
which depends on 
a dimensionless ``load distribution factor'' $\theta$ \cite{Kolo2001}.
The load distribution factor $\theta \in [0,1]$
determines whether the on- or off-rates are affected
by external force, while keeping the ratio of overall on- and off-rate 
unaffected. 
Under force
both the tubulin on-rate $\omega_{\text{on}}$ and the tubulin off-rate
$\omega_{\text{off}}$ now acquire an additional Boltzmann-like  factor. 
 For $\theta=1$, we obtain again $v_{+}(f)$ as given by eq.\
(\ref{eq:vel_f}). The dimensionless stall force is unaffected by $\theta$ and
is still given by 
$f_{\text{stall}}=\ln\left(\omega_{\text{on}}/\omega_{\text{off}} \right)$.

\subsection{Vanishing rescue rate}

We use the generalized force-velocity relation $v_{+}(f,\theta)$ 
given by eq.\ (\ref{eq:vel_theta}) and the 
catastrophe rate $\omega_{c,\text{Flyv}}(f)$ in order to 
calculate the average maximal polymerization force 
$\langle f_{\text{max}} \rangle$ from  the self-consistent 
mean field  eq.\ (\ref{eq:w_r0_f_max}). 
In Fig.\ \ref{fig:vel_theta} (a),
$\langle f_{\text{max}} \rangle$ is shown  as a function of 
the load distribution factor $\theta$ for
$k=10^{-5}\,\rm{N/m}$ and different values of $\omega_{\text{on}}$. At
$\theta=1$, the maximal force 
$\langle f_{\text{max}} \rangle$ equals the maximal polymerization
force obtained with $v_{+}(f)$ from eq.\ (\ref{eq:vel_f}). With decreasing
$\theta$, $\langle f_{\text{max}} \rangle$ increases but remains below
the dimensionless stall force. The growth velocity  $v_{+}(f,\theta)$
increases with decreasing $\theta$ for a fixed force $f$ and, therefore, the
maximal polymerization force $\langle f_{\text{max}} \rangle$ increases.
For high tubulin on-rates $\omega_{\text{on}}=75-100\,
\rm{s}^{-1}$ and small $\theta \approx 0,\ldots,0.2$, 
the maximal polymerization force $\langle f_{\text{max}}\rangle$ 
approaches the dimensionless stall force.

\subsection{Non-zero  rescue rate}

\begin{figure}
 \includegraphics[width=\columnwidth]{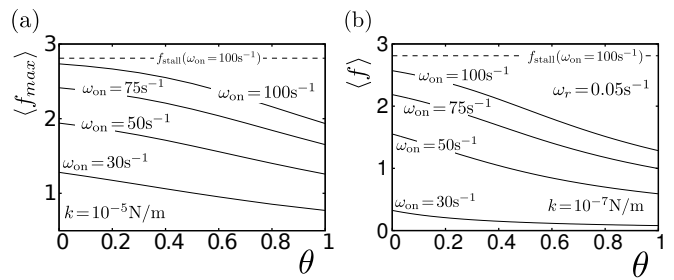}
\caption{ 
   (a): Solid lines: Average maximal polymerization force $\langle
  f_{\text{max}} \rangle$ as a function of $\theta$ for $k=10^{-5}\,\rm{N/m}$
  and different values of $\omega_{\text{on}}$. Dashed line: Dimensionless
  stall force $f_{\text{stall}}$ for $\omega_{\text{on}}=100\,\rm{s}^{-1}$.
  (b): Solid lines: Average steady state force $\langle f \rangle$ as a
  function of $\theta$ for $k=10^{-7}\,\rm{N/m}$,
  $\omega_{r}=0.05\,\rm{s}^{-1}$ and different values of
  $\omega_{\text{on}}$. Dashed line: Dimensionless stall force
  $f_{\text{stall}}$ for
  $\omega_{\text{on}}=100\,\rm{s}^{-1}$. \label{fig:vel_theta}}
\end{figure}

For non-zero rescue rate, the 
 average steady state force $\langle f \rangle $ is calculated 
from the mean field  eq.\ (\ref{eq:mf_fs}), where we use 
the  force-velocity relation
$v_{+}(f,\theta)$ (eq.\ \ref{eq:vel_theta}) and the catastrophe rate
$\omega_{c,\text{Flyv}}(f)$. In Fig.\ \ref{fig:vel_theta} (b), results 
for $\langle f\rangle$ are shown 
 as a function of $\theta$ for $k=10^{-7}\,\rm{N/m}$,
$\omega_{r}=0.05\, \rm{s}^{-1}$ and different values of
$\omega_{\text{on}}$. At $\theta=1$, $\langle f\rangle$ equals the average
steady state force obtained with a velocity 
$v_{+}(f)$ taken from eq.\ (\ref{eq:vel_f}). The
average steady state force $\langle f\rangle$ increases with decreasing
$\theta$, as explained above. For high tubulin on-rates
$\omega_{\text{on}}=75-100\,\rm{s}^{-1}$ and small $\theta
\approx 0,\ldots,0.2$ , also the average steady state force
 $\langle f \rangle$ again approaches the dimensionless
stall force but remains smaller.

\section{Discussion and conclusion}\label{sec:discuss}

We studied MT dynamics in three different confining scenarios:
(i)  confinement by fixed rigid walls, 
(ii) an open system under constant force, 
and (iii) MT growth against an elastic obstacle with a force that depends 
linearly on MT length. These three scenarios represent generic confinement 
scenarios in living cells or geometries, which can be realized experimentally
 {\it in vitro}. 
For all three scenarios, we are able to quantify the MT length distributions. 
In scenario (iii) of an elastic obstacle, stochastic MT growth also 
gives rise to a stochastic force. For this model, we also  quantify 
the average polymerization force generated by the MT in the presence 
of the dynamic instability. 

The 
parameter $\lambda$, see (\ref{eq:lambda}) and (\ref{eq:lambdaf}), governs 
the MT length distributions in confinement by fixed rigid walls, and under 
a constant force. 
For confinement by rigid walls 
we introduced a realistic model for wall-induced catastrophes. 
There is a transition from exponentially increasing to exponentially decreasing 
length distributions if $\lambda$ changes sign. 
The average MT length is increasing for increasing on-rate 
and increasing rescue rate, as shown in 
Figs.\ \ref{fig:conf_x}. 
Wall-induced catastrophes  lead to an overall increase in the 
average catastrophe frequency, which we quantify within the model.

For MT growth under a constant force, there exists a transition between 
bounded and unbounded growth as in the absence of force. This transition 
takes place where the parameter $\lambda(f)$  changes sign. Under force, 
the transition to unbounded growth is shifted to higher on-rates or 
higher rescue rates and determines a critical force $f_c$, 
see Figs.\ \ref{fig:Fconst}.

MT growth under a MT length-dependent linear elastic force allows for 
regulation of the generated polymerization force by experimentally
accessible parameters such as the on-rate or the rescue rate. 
The force is no longer fixed but a stochastically fluctuating quantity
because the MT length is a stochastic quantity. 
 For   zero rescue rate, i.e., in the absence of rescue events, 
we find that the average maximal  polymerization force 
$\langle f_{\text{max}}  \rangle$ before a catastrophe depends
   logarithmically on the tubulin concentration and is always smaller than the
 stall force in the absence of dynamic instability
as shown in Fig.\ \ref{fig:N1_FMAX}.

For a  non-zero
   rescue rate,  we find a  steady state 
   length distribution, which becomes increasingly sharply peaked for 
increasing rescue rate and  is tightly controlled by 
     microtubule growth parameters, see Figs.\ \ref{fig:P(x)}.
Interestingly, the  average microtubule length 
  self-organizes such that the  average steady state polymerization 
force $\langle f \rangle$ equals the critical force for 
  the boundary of bounded and unbounded growth, $\langle f\rangle = f_c$. 
 Because of the sharply peaked MT length distribution, the average 
polymerization force $\langle f \rangle$ can be calculated rather 
accurately within a mean field approach as can be seen in Figs.\ 
\ref{fig:avg_fs} and \ref{fig:avg_fs_k1e05}.
The average polymerization force is always smaller than the
 stall force in the absence of dynamic instability.

Within this mean field approach, we can also describe 
the dynamics of the average 
force, see Figs.\ \ref{fig:mf_time_evo}. This might be useful
in modeling dilution experiments, where the response to sudden changes 
in the on-rate is probed.  For this type of experiment, we estimate 
typical polymerization force relaxation times. 

Finally, we show that our findings are robust against changes 
of the catastrophe model (Figs.\ \ref{fig:exp_wcat_f}) 
as long as the catastrophe rate 
increases exponentially above a characteristic force and that 
results are also robust against variations of the relation 
between force and polymerization velocity in the growing 
phase (Figs.\ \ref{fig:vel_theta}), 
which are obtained by introducing a load distribution factor.

\section{Acknowledgments}

We acknowledge financial support by the Deutsche Forschungsgemeinschaft
(KI 662/4-1).

\appendix
\section{Literature values for parameters}
\begin{table*}
\begin{center}
 
 \begin{tabular}{|c|c|c|c|c|}
\hline
\hline
Ref.\ & $v_{+}(0)$ [m/s] & $\omega_{\text{on}}$ [1/s] &  $v_{-}$ [m/s] & $\omega_{r}$ [1/s]\\
\hline

Drechsel \cite{Drechsel1992} & $(0.7\;...\;2)\cdot 10^{-8}$ & $(11\;...\;32)$ & $\sim 1.8\cdot 10^{-7}$ & - \\

Gildersleeve \cite{Gildersleeve1992} & $\sim 4.2\cdot 10^{-8}$ & $\sim 68$ & $\sim 4.2\cdot 10^{-7}$ & -\\

Walker \cite{Walker1988} & $ (4\;...\;8)\cdot 10^{-8}$ & $(63\;...\;130)$ & $\sim 5\cdot 10^{-7}$ & $(0.05\;...\;0.08) $(TUB)\\

Laan \cite{Laan2008} & $\sim 4.2\cdot 10^{-8}$ & $68.25$ & - & - \\ 

Janson \cite{Janson2004} & $(3\;...\;4.3)\cdot 10^{-8}$ & $(53\;...\;74)$ & - & - \\

Pryer \cite{Pryer1992} & - & - & - & $...\;0.5$ (TUB)  $..\,0.15$ (MAPS) \\

Dhamodharan \cite{Dhamodharan1995} & - & - & - & $ ...\;0.07$ (Cell) $\quad...\;0.085$ (MAPS) \\

Nakao \cite{Nakao2004} & - & - & - & $...\;0.1$ (TUB) \\

Shelden \cite{Shelden1993} & - & - & - & $(0.03\;...\;0.2)$ (Cell) \\
\hline
\hline
\end{tabular}
\caption{Literature values for parameters.
  TUB: {\it in vitro} results for tubulin solutions, Cell: {\it in vivo}
  results,  MAPS: effect from MT associated proteins. 
  Values for $\omega_{\text{on}}$
 are estimated from measured growth velocities via $\omega_{\text{on}} \approx
 v_{+}(0)N/d$ neglecting $\omega_{\text{off}}=6\,{\rm s}^{-1}$ \cite{Janson2004}. 
Here $N=13$ denotes the number of protofilaments within a single MT.}
\label{tab:parameters}
\end{center}

\end{table*}

\begin{table*}
\begin{center}
 \begin{tabular}{|c|c|c|c|c|c|c|c|}
\hline
\hline
 Parameter & $v_{-}\;[\rm{m/s}]$ (see Table \ref{tab:parameters}) & $\omega_{\text{off}}\;[\rm{s}^{-1}]$ &  $d\;[\rm{m}]$ & $r \;[\rm{m}^{-1}\rm{s}^{-1}]$ & $v_{h}\;[\rm{m/s}]$ & $\Delta t\;[\rm{s}]$\\
\hline
Value& $3\cdot 10^{-7}$ & $6$ & $0.6\cdot 10^{-9}$ & $3.7\cdot 10^{6}$ & $4.2\cdot 10^{-9}$ & $0.1$\\
\hline
\hline
\end{tabular}
\caption{Fixed parameter values for calculations and simulations.}
\label{tab:fixed_parameters}
\end{center}
\end{table*}

\end{document}